\documentclass[12pt,draftcls,onecolumn]{IEEEtran}                       % paper
\IEEEoverridecommandlockouts                              % This command is only
\usepackage{graphics} % for pdf, bitmapped graphics files
\usepackage{epsfig} % for postscript graphics files
\usepackage{mathptmx} % assumes new font selection scheme installed
\usepackage{times} % assumes new font selection scheme installed
\usepackage{amsmath} % assumes amsmath package installed
\usepackage{amssymb}  % assumes amsmath package installed
\usepackage{floatrow}
\usepackage{multirow}
\usepackage{algorithm}
\usepackage{algpseudocode}
\usepackage{color}
\usepackage[colorlinks,
           linkcolor=red,
           anchorcolor=green,
           citecolor=blue
           ]{hyperref}
\newtheorem{theorem}{\indent Theorem}

\newtheorem{definition}[theorem]{\indent Definition}

\newtheorem{remark}{\indent Remark}

\title{\LARGE \bf
Quantum Ensemble Classification: A Sampling-based Learning Control
Approach}
\author{Chunlin~Chen,
        Daoyi~Dong,
        Bo Qi,
        Ian R.~Petersen,
        Herschel~Rabitz
\thanks{This work was supported by the Natural Science
Foundation of China (Nos.61273327 and 61004049), by the Australian Research
Council (DP130101658 and FL110100020) and by the US NSF (No. CHE-0718610).}
\thanks{C. Chen is with the Department of Control and System Engineering, Nanjing
University, Nanjing 210093, China and with the Department of
Chemistry, Princeton University, Princeton, New Jersey 08544, USA
(email: clchen@nju.edu.cn).}
\thanks{D. Dong is with the School of Engineering and Information
Technology, University of New South Wales at the Australian Defence
Force Academy, Canberra, ACT 2600, Australia (email:
daoyidong@gmail.com).}
\thanks{B. Qi is with the Key Laboratory of Systems and Control, Academy of
Mathematics and Systems Science, Chinese Academy of Sciences,
Beijing 100190, China (email: qibo@amss.ac.cn).}
\thanks{I. R. Petersen is with the School of Engineering and Information
Technology, University of New South Wales at the Australian
Defence Force Academy, Canberra, ACT 2600, Australia (email:
i.r.petersen@gmail.com).}
\thanks{H. Rabitz is with the Department of
Chemistry, Princeton University, Princeton, New Jersey 08544, USA
(email: hrabitz@princeton.edu).}}

\begin{document}

\maketitle
%\thispagestyle{empty}  %\pagestyle{empty}
%%%%%%%%%%%%%%%%%%%%%%%%%%%%%%%%%%%%%%%%%%%%%%%%%%%%%%%%%%%%%%%%%%%%%%%%%%%%%%%%
\begin{abstract}
Quantum ensemble classification has significant
applications in discrimination of atoms (or molecules), separation
of isotopic molecules and quantum information extraction.
However, quantum mechanics forbids deterministic discrimination
among nonorthogonal states. The classification of inhomogeneous
quantum ensembles is very challenging since there exist variations in the parameters
characterizing the members within
different classes. In this paper, we recast quantum ensemble
classification as a supervised quantum learning problem. A systematic classification methodology is presented
by using a sampling-based learning control (SLC)
approach for quantum discrimination. The classification task is
accomplished via simultaneously steering members belonging to different classes to their corresponding target states
(e.g., mutually orthogonal states). Firstly a new discrimination method is proposed for
two similar quantum systems. Then an SLC method is
presented for quantum ensemble classification. Numerical results demonstrate the effectiveness of the proposed approach for the binary
classification of two-level quantum ensembles and the
multiclass classification of multilevel quantum ensembles.
\end{abstract}


%%%%%%%%%%%%%%%%%%%%%%%%%%%%%%%%%%%%%%%%%%%%%%%%%%%%%%%%%%%%%%%%%%%%%%%%%%%%%%%%
\section{Introduction}\label{Sec1}
Optimal discrimination \cite{Mohseni et al 2004}, \cite{Beltrani et
al 2009} and classification \cite{Guta and Kotlowski 2010} of
quantum states or quantum systems is a central topic in quantum
information technology \cite{Nielsen and Chuang 2000}. This interdisciplinary research area involves pattern recognition in machine learning, quantum control in quantum technology \cite{Altafini and Ticozzi 2012}-\cite{Wiseman and Milburn 2009} and very common laboratory problems of isolating
similar species in chemical physics. In existing research, discrimination of two similar quantum
systems (e.g., similar molecules) has been extensively investigated
\cite{Li et al 2002}-\cite{Chakrabarti et al 2008Pareto}.

Many practical quantum systems exist in the form of quantum ensembles.
A quantum ensemble consists of a large number of (e.g., $10^{23}$)
single quantum systems (e.g., identical spin systems or molecules). Each single quantum system
in a quantum ensemble is referred to as a member of the ensemble.
Quantum ensembles have wide applications in emerging quantum
technology including quantum computation \cite{Cory et al 1997},
long-distance quantum communication \cite{Duan et 2001}, and magnetic resonance imaging
\cite{Li et al 2011PNAS}. In practical
applications, the members of a quantum ensemble could show
variations in the parameters that characterize the system dynamics
\cite{Levitt 1986}, \cite{Li and Khaneja 2006}. Such an ensemble is called an
inhomogeneous quantum ensemble \cite{Chen et al 2013}. For example, the spins of an ensemble in nuclear magnetic
resonance (NMR) experiments may have a large dispersion in the
strength of the applied radio frequency field (RF imhomogeneity) and
their natural frequencies (Larmor dispersion) \cite{Li and Khaneja
2006}. In complex systems, there are intrinsic inhomogeneities of even chemically identical
molecules due to different conformations and diverse environments
\cite{Brinks et al 2010}.
The classification of inhomogeneous quantum ensembles is a significant
issue and has great potential applications in the discrimination of
atoms (or molecules), the separation of isotopic molecules and quantum
information extraction.

However, quantum mechanics forbids deterministic discrimination among
nonorthogonal states \cite{Mohseni et al 2004}. A useful idea is to first drive the
members of a quantum ensemble from an initial state to different
orthogonal states corresponding to different classes (e.g., eigenstates) before classifying
them. Usually, it is impractical to employ different control
inputs for individual members of a quantum ensemble in physical
experiments. Hence, it is important to develop new approaches for
designing external control fields that can simultaneously steer the
ensemble of inhomogeneous systems from an initial state to different
target states when variations exist in their internal parameters.
Some quantum control techniques such as the multidimensional pseudospectral method
\cite{Li et al 2011PNAS}, \cite{Ruths and Li 2011}, the Lyapunov control methodology \cite{Beauchard et al 2012} and the sampling-based learning control approach \cite{Chen et al 2013} may provide inspiration for the solution to this problem.

In this paper, we recast the quantum ensemble classification task as
a supervised quantum learning problem and present a systematic
classification methodology by using a sampling-based learning control (SLC) method \cite{Chen et al 2013}, \cite{Dong et al 2013}  in quantum
discrimination. In this method, we first learn an optimal control strategy to steer the members in a quantum ensemble
belonging to different classes into their corresponding target states, and then employ a physical read-out process (e.g., projective
measurement, fluorescence images of molecules
\cite{Brinks et al 2010}, Stern-Gerlach experiment for spin
systems \cite{Nielsen and Chuang 2000}, \cite{Appleby 2000}) to classify these classes. For example, the states of single molecules could be
read out using a visualization technique, where the highly photostable chromophore
dinaphtoquaterrylenebis (dicarboximide) (DNQDI) is embedded in thin
polymer films in concentrations sufficiently low to allow individual
DNQDI molecules to be spatially resolved in an epifluorescence
confocal microscope (for details, see \cite{Brinks et al 2010}). It is feasible to read out the intensity of the
single-molecule fluorescence after they are excited with laser
pulses. For a spin ensemble, when some members are driven to spin up and the other are driven to spin down, it
is feasible to physically separate the two classes of members using Stern-Gerlach experiments \cite{Nielsen and Chuang 2000}.
We first develop a new approach for the discrimination of two similar quantum systems and the binary classification of quantum ensembles. Then
we apply the proposed approach to multiclass classification of multilevel quantum ensembles.

This paper is organized as follows. Section \ref{Sec2}
formulates the learning problem for quantum ensemble classification.
A control design method is presented in Section \ref{Sec3} for
quantum discrimination of similar quantum systems. In Section \ref{Sec4} an
SLC method is proposed for the binary classification of quantum ensembles and
numerical results are demonstrated for an
ensemble of two-level spin systems. The proposed approach is applied to
the multiclass classification of multi-level quantum ensembles in Section \ref{Sec5}. Conclusions are presented
in Section \ref{Sec6}.

\section{Problem formulation}\label{Sec2}
We focus on finite-dimensional closed quantum systems. The
state evolution of a quantum system is described by the
following Schr\"{o}dinger equation (setting the reduced Plank
constant $\hbar=1$):
\begin{equation} \label{systemmodel}
\left\{ \begin{array}{l}
  \frac{d}{dt}|{\psi}(t)\rangle=-iH(t)|\psi(t)\rangle \\
 t\in [0, T], \ |\psi(0)\rangle=|\psi_{0}\rangle\\
\end{array}
\right.
\end{equation}
where $|\psi(t)\rangle$ (quantum state) is a unit complex vector on the underlying Hilbert space, $H(t)$ is the system
Hamiltonian and $i=\sqrt{-1}$.
The dynamics of the system is governed by a time-dependent
Hamiltonian of the form
\begin{equation}\label{Hamiltonian}
H(t)=H_{0}+H_{c}(t)=H_{0}+\sum_{m=1}^{M}u_{m}(t)H_{m},
\end{equation}
where $H_{0}$ is the free Hamiltonian of the system and
$H_{c}(t)=\sum_{m=1}^{M}u_{m}(t)H_{m}$ is the time-dependent control
Hamiltonian that represents the interaction of the system with the
external fields $u_{m}(t)$ (real-valued and square-integrable
functions). $H_{m}$ are Hermitian operators through which the
controls couple to the system. The solution of
(\ref{systemmodel}) is given by $\displaystyle
|\psi(t)\rangle=U(t)|\psi_{0}\rangle$, where the propagator $U(t)$
satisfies the following equation ($I$ is an identity matrix)
\begin{equation}
\left\{ \begin{array}{c}
  \frac{d}{dt}U(t)=-iH(t)U(t),\\
  t\in [0, T], \ U(0)=I.\\
\end{array}
\right.
\end{equation}

In this paper, we consider the classification problem for a quantum ensemble of similar members with different
Hamiltonians, which is referred to as
\emph{quantum ensemble classification} (QEC). Suppose that for an inhomogeneous quantum ensemble,
we are given an unknown member belonging to a certain class, how
well can we predict the class that the unknown member belongs to? In classical
machine learning, this problem can be solved using typical supervised
learning algorithms with a training set. However, this problem
is much more difficult for quantum systems because we can not achieve deterministic
discrimination for given quantum systems unless they lie in mutually orthogonal
states. We have to drive the members from different classes to appropriate
orthogonal states (e.g., eigenstates) before we can discriminate
them with high accuracy. The sampling-based learning control approach presented for the control of inhomogeneous quantum ensembles can be combined with supervised learning for QEC. We define the training set
for the QEC problem as follows.

\begin{definition}[Training set of QEC]\label{TrainingSet}
A training set consists of $N$ quantum systems (each of them labeled
with an associated class) that are chosen from the quantum ensemble
and the set is denoted as
\begin{equation}
D_N=\{(H^1(t),y_1),(H^2(t),y_2),\ldots,(H^N(t),y_N)\}
\end{equation}
where $H^n(t)$ ($n=1,2,\ldots, N$) describes the $n\text{th}$ quantum
system in the training set and $y_n$ is the associated class that
this quantum system belongs to.
\end{definition}

For ease of presentation, we first consider an inhomogeneous
ensemble consisting of two classes of members (i.e., classes \emph{A}
and \emph{B}) and propose an SLC approach for this binary quantum
ensemble classification problem using a spin-$\frac{1}{2}$ quantum
ensemble example. We further extend the proposed
approach to the classification problem with multi classes and multi-level
quantum ensembles. For the binary quantum ensemble classification
problem, the Hamiltonian of each member has the following form
\begin{equation}\label{classAB}
\left \{
\begin{split} & H^A_{\varepsilon_0,
\varepsilon_u}(t)=g^A_0(\varepsilon_0)H_{0}+g^A_u(\varepsilon_u)\sum_{m=1}^{M}u_{m}(t)H_{m}\\
&  H^B_{\varepsilon_0,
\varepsilon_u}(t)=g^B_0(\varepsilon_0)H_{0}+g^B_u(\varepsilon_u)\sum_{m=1}^{M}u_{m}(t)H_{m}.
\end{split}
\right.
\end{equation}
$g^A_0(\cdot)$ and $g^A_u(\cdot)$ are known functions, while the inhomogeneity parameters $\varepsilon_0$ and $ \varepsilon_u$  in the Hamiltonian
$H^A_{\varepsilon_0, \varepsilon_u}(t)$ for class \emph{A} are characterized by the distribution functions $d^A_0(\varepsilon_0)$ and $d^A_u(\varepsilon_u)$, respectively. We
assume that the parameters $\varepsilon_0$ and $\varepsilon_u$  are
time independent. A similar expression to (\ref{classAB}) is defined
for the Hamiltonian $H^B_{\varepsilon_0, \varepsilon_u}(t)$ of class
\emph{B}.

\begin{figure}
\centering
\includegraphics[width=4.6in]{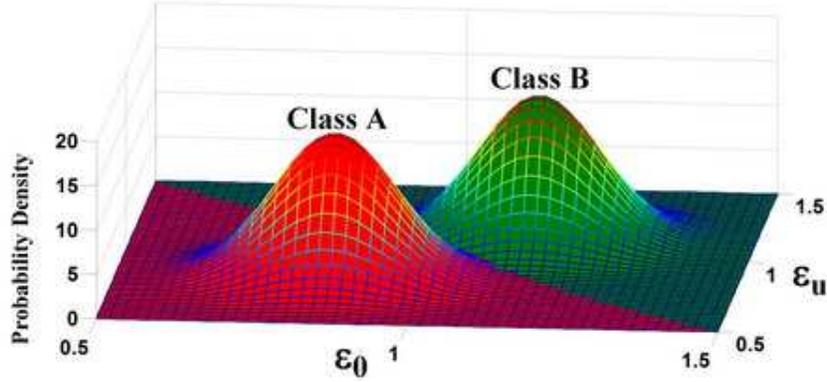}
\caption{Example of an inhomogeneous quantum ensemble consisting
of two classes (\emph{A} and \emph{B}) with the inhomogeneity parameters having Gaussian
distribution.}
\end{figure}

An example is shown in Fig. 1 to describe the inhomogeneity of the
quantum ensemble consisting of two classes of members
with parameters in each class having Gaussian distribution. The
function $d^A_0(\varepsilon_0)$ ($d^B_0(\varepsilon_0)$)
characterizes the distribution of inhomogeneity in the free
Hamiltonian for class \emph{A} (\emph{B}) and $d^A_u(\varepsilon_u)$
($d^B_u(\varepsilon_u)$) characterizes the distribution of
inhomogeneity in the control Hamiltonian for class \emph{A}
(\emph{B}). Fig. 1 shows a 2D Gaussian distribution case regarding
the parameters $\varepsilon_0$ and $\varepsilon_u$.

For a binary quantum ensemble classification task, the objective
is to design a control strategy $u(t)=\{u_{m}(t), m=1,2,\ldots ,
M\}$ to simultaneously stabilize the members in class \emph{A}
(with different $\varepsilon_0$ and $\varepsilon_u$) from an initial
state $|\psi_{0}\rangle$ to the same target state
$|\psi_{\text{targetA}}\rangle$, and at the same
time to stabilize the members in class \emph{B} (with different
$\varepsilon_0$ and $\varepsilon_u$) from
$|\psi_{0}\rangle$ to another target state
$|\psi_{\text{targetB}}\rangle$. A binary QEC problem can be described by the following definition.

\begin{definition}[Binary QEC]\label{BQEC}
A binary quantum ensemble classification (binary QEC) task is to
construct a binary quantum classifier to maximize the classification
accuracy, where this binary quantum classifier consists of three
steps:

\begin{enumerate}
 \item \emph{Training step:} Learn an optimal control strategy $u(t)$ with the training
 set
 $$D_N=\{(H^1(t),y_1),(H^2(t),y_2),\ldots,(H^N(t),y_N)\},$$
 where
 $y_n \in \{A, B\}$ ($A$ and $B$ are symbolic constants) and $H^n(t)$ ($n=1,2,\ldots,N$) is the time-dependent Hamiltonian
 describing the $n\text{th}$ member in the training set.

 \item \emph{Coherent control step:} Apply the learned optimal control strategy
 $u(t)$ to all the members of the quantum ensemble.

 \item \emph{Classification step:} Predict the
 class $y_j$ of an unknown quantum system in the quantum ensemble using a corresponding physical read-out process,
 where $j=1,2,\ldots, N_e$ and $N_e$ is the number of members in the quantum ensemble.
\end{enumerate}
\end{definition}

For example, a schematic of the classification process for a spin-$\frac{1}{2}$ quantum ensemble is
demonstrated in Fig. 2.
As shown in Fig. 2, an ensemble of inhomogeneous spin-$\frac{1}{2}$
systems is prepared with an initial state. After learning using a
training set from the quantum ensemble, we can find an optimal
control strategy to simultaneously drive all the members of class
\emph{A} to the target state (spin up) and all the members of
class \emph{B} to another target state (spin down). Then we can use a Stern-Gerlach experiment to physically separate the two classes.

\begin{figure}
\centering
\includegraphics[width=4.5in]{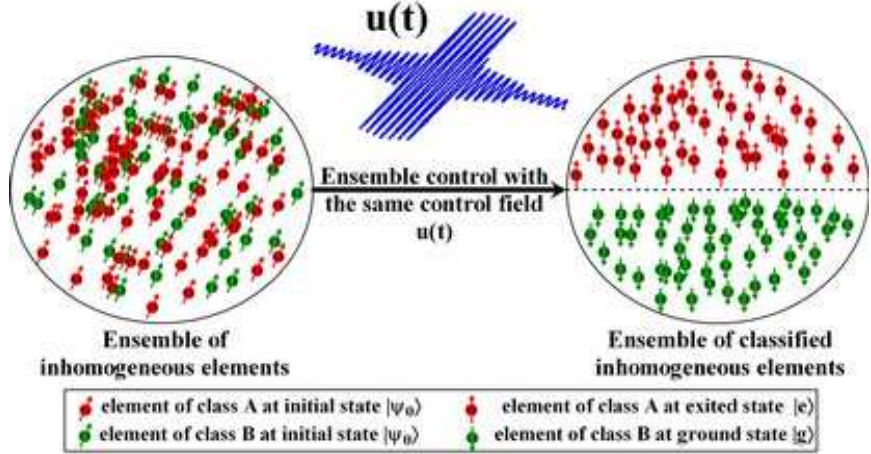}
\caption{Schematic of the binary classification for a
spin-$\frac{1}{2}$ quantum ensemble.}
\end{figure}

From Definition \ref{BQEC} it is clear that the key task of the classification problem is to learn an optimal control strategy in the training
step for the binary
quantum classifier. The training performance is described by a
\emph{performance function} $J(u)$ for each learned control strategy
$u=\{u_{m}(t), m=1,2,\ldots , M\}$. The binary QEC problem can then
be formulated as a maximization problem as follows:
\begin{equation}\label{ensemble control}
\begin{split}
\displaystyle  \ \ \  & \max_u
J(u):=\max_{u} \{w_A\mathbb{E}[J^A_{\varepsilon_0,\varepsilon_u}(u)]+w_B\mathbb{E}[J^B_{\varepsilon_0,\varepsilon_u}(u)]\}\\
\text{s.t.} \ \ \ & t \in [0, T]\\
&|\psi^A_{\varepsilon_0,\varepsilon_u}(0)\rangle=|\psi^B_{\varepsilon_0,\varepsilon_u}(0)\rangle=|\psi_{0}\rangle \\
&\left \{
\begin{split}
&\frac{d}{dt}|\psi^A_{\varepsilon_0,\varepsilon_u}(t)\rangle=-iH^A_{\varepsilon_0,\varepsilon_u}(t)|\psi^A_{\varepsilon_0,\varepsilon_u}(t)\rangle \\
& H^A_{\varepsilon_0,
\varepsilon_u}(t)=g^A_0(\varepsilon_0)H_{0}+g^A_u(\varepsilon_u)\sum_{m=1}^{M}u_{m}(t)H_{m}\\
&J^A_{\varepsilon_0,\varepsilon_u}(u):=\vert
\langle\psi^A_{\varepsilon_0,\varepsilon_u}(T)|\psi_{\text{targetA}}\rangle\vert^{2}\\
\end{split} \right.\\
&\left \{
\begin{split}
&\frac{d}{dt}|\psi^B_{\varepsilon_0,\varepsilon_u}(t)\rangle=-iH^B_{\varepsilon_0,\varepsilon_u}(t)|\psi^B_{\varepsilon_0,\varepsilon_u}(t)\rangle \\
& H^B_{\varepsilon_0,
\varepsilon_u}(t)=g^B_0(\varepsilon_0)H_{0}+g^B_u(\varepsilon_u)\sum_{m=1}^{M}u_{m}(t)H_{m}\\
&J^B_{\varepsilon_0,\varepsilon_u}(u):=\vert
\langle\psi^B_{\varepsilon_0,\varepsilon_u}(T)|\psi_{\text{targetB}}\rangle\vert^{2}\\
\end{split}\right.\\
\end{split}
\end{equation}
where $w_A, w_B \in [0,1]$ are the weights assigned to classes
\emph{A} and \emph{B}, respectively, satisfying $w_A+w_B=1$.
$J^A_{\varepsilon_0,\varepsilon_u}(u)$ is a measure of
classification accuracy for each member in class \emph{A} regarding
the target state $|\psi_{\text{targetA}}\rangle$ and
$\mathbb{E}[J^A_{\varepsilon_0,\varepsilon_u}(u)]$ denotes the
average value of $J^A_{\varepsilon_0,\varepsilon_u}(u)$ over class
\emph{A}. A similar expression holds for class \emph{B}. It is
clear that $J^A_{\varepsilon_0,\varepsilon_u}$ and
$J^B_{\varepsilon_0,\varepsilon_u}$ depend implicitly on the control
strategy $u(t)$ through the Schr\"odinger equation. The performance $J(u)$ represents
the weighted accuracy of classification.

\section{Discrimination of two similar quantum systems}\label{Sec3}
Optimal dynamic discrimination between two similar quantum systems has been investigated using different techniques
\cite{Beltrani et al 2009}, \cite{Li et al 2002}. The quantum discrimination problem can be taken as a special case of the binary QEC problem with the number of members in an ensemble
$N_{e}=2$. In this sense, control design for
quantum discrimination is the foundation of quantum ensemble
classification. In this section, we develop a gradient-based
learning control method for quantum discrimination of two similar
quantum systems and then we extend the method to control
design for binary QEC in Section \ref{Sec4}.

\subsection{Learning control design for quantum discrimination}\label{Subsec3.1}
Suppose two similar quantum systems to be discriminated $a$ and
$b$ have the following Hamiltonians:
\begin{equation}\label{discriminationAB}
\left \{
\begin{split} & H^a_{\varepsilon^a_0,\varepsilon^a_u}(t)=g_0(\varepsilon^a_0)H_{0}+g_u(\varepsilon^a_u)\sum_{m=1}^{M}u_{m}(t)H_{m}\\
&
H^b_{\varepsilon^b_0,\varepsilon^b_u}(t)=g_0(\varepsilon^b_0)H_{0}+g_u(\varepsilon^b_u)\sum_{m=1}^{M}u_{m}(t)H_{m}
\end{split}
\right.
\end{equation}
where $\varepsilon^a_0$, $\varepsilon^a_u$, $\varepsilon^b_0$ and
$\varepsilon^b_u$ are predefined constants for functions $g_0(\cdot)$ and
$g_u(\cdot)$. \emph{a} and \emph{b} are prepared in
the same initial state $|\psi_0\rangle$. The objective is to find an
optimal control strategy $u(t)$ ($t\in [0, T]$) to
drive the state of system \emph{a} to the target state
$|\psi_{\text{targetA}}\rangle$ and the state of system \emph{b} to
the target state $|\psi_{\text{targetB}}\rangle$ at the same time. Usually, we let $\langle \psi_{\text{targetA}}|\psi_{\text{targetB}}\rangle=0$ so that we can
completely discriminate system \emph{a} from system \emph{b}. The control performance $J(u)$
is redefined for the discrimination problem as
\begin{equation}
J(u):=w_aJ^a_{\varepsilon^a_0,\varepsilon^a_u}(u)+w_bJ^b_{\varepsilon^b_0,\varepsilon^b_u}(u)
\end{equation}
where $w_a, w_b \in [0, 1]$ are the weights assigned to the
associated systems, respectively, and
\begin{equation}
\begin{split}
& J^a_{\varepsilon^a_0,\varepsilon^a_u}(u):=\vert
\langle\psi^a_{\varepsilon^a_0,\varepsilon^a_u}(T)|\psi_{\text{targetA}}\rangle\vert^{2},\\
& J^b_{\varepsilon^b_0,\varepsilon^b_u}(u):=\vert
\langle\psi^b_{\varepsilon^b_0,\varepsilon^b_u}(T)|\psi_{\text{targetB}}\rangle\vert^{2}.\\
\end{split}
\end{equation}
Here we set $w_a=w_b=0.5$ for the discrimination problem.

In order to find an optimal control strategy
$u^{*}=\{u^{*}_{m}(t), (t \in [0,T]), m=1,2,\ldots, M\}$ for the
discrimination problem, it is a good choice to follow the direction
of the gradient of $J(u)$ as an ascent direction. For ease of
notation, we present the method for $M=1$. We introduce a time-like
variable $s$ to characterize different control strategies
$u^{(s)}(t)$. Then a gradient flow in the control space can be
defined as

\begin{equation}\label{gradientflowequation}
\frac{du^{(s)}}{ds} =\nabla J(u^{(s)}),
\end{equation}
where $\nabla J(u)$ denotes the gradient of $J(u)$ with respect to
the control $u$. It is easy to show that if $u^{(s)}$ is the
solution of \eqref{gradientflowequation} starting from an arbitrary
initial condition $u^{(0)}$, then the value of $J(u)$ is increasing
along $u^{(s)}$, i.e., $\frac{d}{ds}J(u^{(s)})\geq 0$. In other
words, starting from an initial guess $u^{0}$, we solve the
following initial value problem
\begin{equation}\label{gradientflowequation2}
\left\{%
\begin{split}
  & \frac{du^{(s)}}{ds} = \nabla J(u^{(s)}) = w_a\nabla J^a_{\varepsilon^a_0,\varepsilon^a_u}(u^{(s)})+w_b\nabla J^b_{\varepsilon^b_0,\varepsilon^b_u}(u^{(s)}) \\
  & u^{(0)}=u^{0} \\
\end{split}%
\right.
\end{equation}
in order to find a control strategy which maximizes $J(u)$. This
initial value problem can then be solved numerically by a forward
Euler method (or other high order integration methods) over the
$s$-domain, i.e.,
\begin{equation}\label{iteration1}
u(s+\triangle s, t)=u(s,t)+\triangle s\nabla J(u^{(s)}).
\end{equation}

As for practical applications, we present its iterative
approximation version to find the optimal control $u^*(t)$, where we use $k$ as an index of iterations
instead of the variable $s$ and denote the control at iteration
step $k$ as $u^{k}(t)$. Equation \eqref{iteration1} can be rewritten
as
\begin{equation}\label{iteration2}
u^{k+1}(t)=u^{k}(t)+ \eta_{k}\nabla J(u^{k}),
\end{equation}
where $\eta_{k}$ is the updating step (learning rate) for the $k\text{th}$ iteration and
\begin{equation}
\nabla J(u^{k}) = w_a\nabla
J^a_{\varepsilon^a_0,\varepsilon^a_u}(u^{k})+w_b\nabla
J^b_{\varepsilon^b_0,\varepsilon^b_u}(u^{k}).
\end{equation}
In addition, we have the gradient of
$J^a_{\varepsilon^a_0,\varepsilon^a_u}(u^k)$ with respect to the
control $u$ as follows (a detailed derivation is provided in the
appendix)
\begin{equation}
\nabla
J^a_{\varepsilon^a_0,\varepsilon^a_u}(u^k)=2\Im\left(\langle\psi^a_{\varepsilon^a_0,\varepsilon^a_u}(T)\vert\psi_{\textrm{targetA}}\rangle\langle\psi_{\textrm{targetA}}\vert
G^a_1(t)\vert\psi_0\rangle\right),
\end{equation}
where $\Im(\cdot)$ denotes the imaginary part of a complex number,
$G^a_1(t)=U_{\varepsilon^a_0,\varepsilon^a_u}(T)U_{\varepsilon^a_0,\varepsilon^a_u}^\dagger(t)g_u(\varepsilon^a_u)H_1U_{\varepsilon^a_0,\varepsilon^a_u}(t)$,
and the propagator $U_{\varepsilon^a_0,\varepsilon^a_u}(t)$
satisfies
$$\frac{d}{dt}U_{\varepsilon^a_0,\varepsilon^a_u}(t)=-iH^a_{\varepsilon^a_0,\varepsilon^a_u}(t)U_{\varepsilon^a_0,\varepsilon^a_u}(t),\quad U(0)=I.$$
A similar expression can also be derived for $\nabla
J^b_{\varepsilon^b_0,\varepsilon^b_u}(u^k)$. When we generalize the
gradient flow method to the case with $M>1$, for each control
$u_m(t)$ ($m=1,2,\ldots, M$) of the control strategy $u(t)$, we have
\begin{equation}\small \label{gradientJ}
\begin{split} \nabla J(u^{k}_m) = &
2w_a\Im\left(\langle\psi^a_{\varepsilon^a_0,\varepsilon^a_u}(T)\vert\psi_{\textrm{targetA}}\rangle\langle\psi_{\textrm{targetA}}\vert
G^a_m(t)\vert\psi_0\rangle\right)\\
&
+2w_b\Im\left(\langle\psi^b_{\varepsilon^b_0,\varepsilon^b_u}(T)\vert\psi_{\textrm{targetB}}\rangle\langle\psi_{\textrm{targetB}}\vert
G^b_m(t)\vert\psi_0\rangle\right)
\end{split}
\end{equation}
where
$$G^a_m(t)=U_{\varepsilon^a_0,\varepsilon^a_u}(T)U_{\varepsilon^a_0,\varepsilon^a_u}^\dagger(t)g_u(\varepsilon^a_u)H_mU_{\varepsilon^a_0,\varepsilon^a_u}(t),$$
$$G^b_m(t)=U_{\varepsilon^b_0,\varepsilon^b_u}(T)U_{\varepsilon^b_0,\varepsilon^b_u}^\dagger(t)g_u(\varepsilon^b_u)H_mU_{\varepsilon^b_0,\varepsilon^b_u}(t).$$

A gradient flow based iterative learning algorithm for the discrimination of quantum
systems is shown in \emph{Algorithm 1}.

\begin{algorithm}
\caption{Gradient flow based iterative learning for quantum
discrimination} \label{ModifiedGradientFlow}
\begin{algorithmic}[1]

\State Set the index of iterations $k=0$ \State Choose a set of
arbitrary controls $u^{k=0}(t)=\{u_{m}^{0}(t),\ m=1,2,\ldots,M\}, t
\in [0,T]$

\Repeat {\ (for each iterative process)}

\State Compute the propagator
$U_{\varepsilon^a_0,\varepsilon^a_u}^{k}(t)$ and
$U_{\varepsilon^b_0,\varepsilon^b_u}^{k}(t)$ for systems \emph{a}
and \emph{b}, respectively, with the control strategy $u^{k}(t)$

\Repeat {\ (for each control $u_{m}(t)$ ($m=1,2,\ldots,M$) of the
control vector $u^{k}(t)$)}

\State $\delta_m^{k}(t):=\nabla J(u^{k}_m)$ and compute $\nabla
J(u^{k}_m)$ using equation \eqref{gradientJ}

\State $u_{m}^{k+1}(t)=u_{m}^{k}(t)+\eta_{k} \delta_{m}^{k}(t)$

\Until {\ $m=M$}

\State $k=k+1$

\Until {\ the learning process ends}

\State The optimal control strategy
$u^{*}(t)=\{u_{m}^*(t)\}=\{u_{m}^{k}(t)\}, \ m=1,2,\ldots,M$

\end{algorithmic}
\end{algorithm}

\begin{remark}
The numerical solution of control design using \emph{Algorithm
1} is always difficult with a time varying continuous control
strategy $u(t)$. In the practical implementation, we usually
divide the time duration $[0,T]$ equally into a number of time
slices $\triangle t$ and assume that the controls are constant
within each time slice. Instead of $t \in [0,T]$, the time index
is $t_{q}=qT/Q$, where $Q=T/\triangle t$ and $q=1,2,\ldots,Q$.
\end{remark}

\subsection{Numerical examples}\label{Subsec3.2}
To demonstrate this learning control method for discrimination of two
similar quantum systems, we consider two-level
(spin-$\frac{1}{2}$) systems. We denote the Pauli matrices
$\sigma=(\sigma_{x},\sigma_{y},\sigma_{z})$ as follows:
\begin{equation}
\sigma_{x}=\begin{pmatrix}
  0 & 1  \\
  1 & 0  \\
\end{pmatrix} , \ \ \ \
\sigma_{y}=\begin{pmatrix}
  0 & -i  \\
  i & 0  \\
\end{pmatrix} , \ \ \ \
\sigma_{z}=\begin{pmatrix}
  1 & 0  \\
  0 & -1  \\
\end{pmatrix} .
\end{equation}
For a two-level quantum system, we may assume the free Hamiltonian
$H_{0}=\frac{1}{2}\sigma_{z}$. Its two eigenstates are denoted as
$|0\rangle$ (e.g., spin up) and $|1\rangle$ (e.g., spin down). To
control a two-level quantum system, we use the control Hamiltonian
of
$H_{u}=\frac{1}{2}u_{1}(t)\sigma_{x}+\frac{1}{2}u_{2}(t)\sigma_{y}$.
Hence,
%\begin{equation}\label{generalmodel}
%|\dot{\psi}(t)\rangle=-iH(t)|\psi(t)\rangle,
%\end{equation}
%where
\begin{equation}H(t)=H_{0}+H_{u}(t)=\frac{1}{2}\sigma_{z}+\frac{1}{2}u_{1}(t)\sigma_{x}+\frac{1}{2}u_{2}(t)\sigma_{y}.\end{equation}
For two similar spin-$\frac{1}{2}$ systems, the Hamiltonian of
each system can be described as
\begin{equation}
\begin{split}
H_{\varepsilon_0,\varepsilon_u}(t)&=g_0(\varepsilon_0)H_{0}+g_u(\varepsilon_u)H_{u}(t)\\
&=\frac{1}{2}g_0(\varepsilon_0)\sigma_{z}+\frac{1}{2}g_u(\varepsilon_u)(u_{1}(t)\sigma_{x}+u_{2}(t)\sigma_{y}).
\end{split}
\end{equation}
We assume $g_0(\varepsilon_0)=\varepsilon_0$ and
$g_u(\varepsilon_u)=\varepsilon_u$. The state of the two
quantum systems can be represented in the eigen-basis of $H_0$ by
$|\psi(t)\rangle=c_{0}(t)|0\rangle+c_{1}(t)|1\rangle$. Denote
$C(t)=(c_{0}(t),c_{1}(t))^{T}$, where $c_0(t)$ and $c_1(t)$ are
complex numbers, and $x^{T}$ represents the transpose of $x$. We have
\begin{equation}\label{ensemble2level}
\left(%
\begin{array}{c}
  \dot{c}_{0}(t) \\
  \dot{c}_{1}(t) \\
\end{array}%
\right)=
\left(%
\begin{array}{cc}
  0.5\varepsilon_0 i & \varepsilon_u f(u) \\
  -\varepsilon_u f^*(u)& -0.5\varepsilon_0 i \\
\end{array}%
\right) \left(%
\begin{array}{c}
  c_{0}(t) \\
  c_{1}(t) \\
\end{array}%
\right),
\end{equation}
where $f(u)=u_{2}(t)-0.5iu_{1}(t)$,
$(\varepsilon_0,\varepsilon_u)=(\varepsilon^a_0,\varepsilon^a_u)$
for system \emph{a} and
$(\varepsilon_0,\varepsilon_u)=(\varepsilon^b_0,\varepsilon^b_u)$
for system \emph{b}.

Define the performance function as
\begin{equation}\label{2level-cost}
\begin{split}
J(u)&=\frac{1}{2}J^a(u)+\frac{1}{2}J^b(u)\\
&=\frac{1}{2}\vert \langle
C_{a}(T)|C_{\text{targetA}}\rangle\vert^{2}+\frac{1}{2}\vert \langle
C_{b}(T)|C_{\text{targetB}}\rangle\vert^{2}.
\end{split}
\end{equation}
The task is to find a control $u(t)$ to maximize the performance
function in (\ref{2level-cost}). For a given small threshold
$\epsilon>0$, if $|J(u^{k+1})-J(u^{k})|<\epsilon$ for
uninterrupted $n_e$ steps, we may think we find a suitable control
law for the problem. In this paper, we set $\epsilon=10^{-4}$ and
$n_e=100$ in all numerical experiments.

Now we employ \emph{Algorithm 1} to find the optimal control
strategy $u^{*}(t)=\{u^{*}_{m}(t), m=1,2\}$ and then apply the
optimal control strategy for discriminating system \emph{a} from
system \emph{b}. The parameter settings are listed as follows: the
initial state $C_{0}=(1,0)^{T}$, i.e.,
$|\psi_{0}\rangle=|0\rangle$, and the target state for system
\emph{a} $C_{\text{targetA}}=(1,0)^{T}$, i.e.,
$|\psi_{\text{targetA}}\rangle=|0\rangle$; the target state for
system \emph{b} $C_{\text{targetB}}=(0,1)^{T}$, i.e.,
$|\psi_{\text{targetB}}\rangle=|1\rangle$; The ending time $T=5$
(in atomic units) and the total time duration $[0,T]$ is equally
discretized into $Q=500$ time slices with each time slice $\Delta
t=(t_{q}-t_{q-1})|_{q=1,2,\ldots,Q}=T/Q=0.01$; the learning rate
$\eta_{k}=0.2$; the control strategy is initialized as
$u^{k=0}(t)=\{u^{0}_{1}(t)=\sin t, u^{0}_{2}(t)=\sin t\}$.

\begin{figure}
\centering
\includegraphics[width=4.5in]{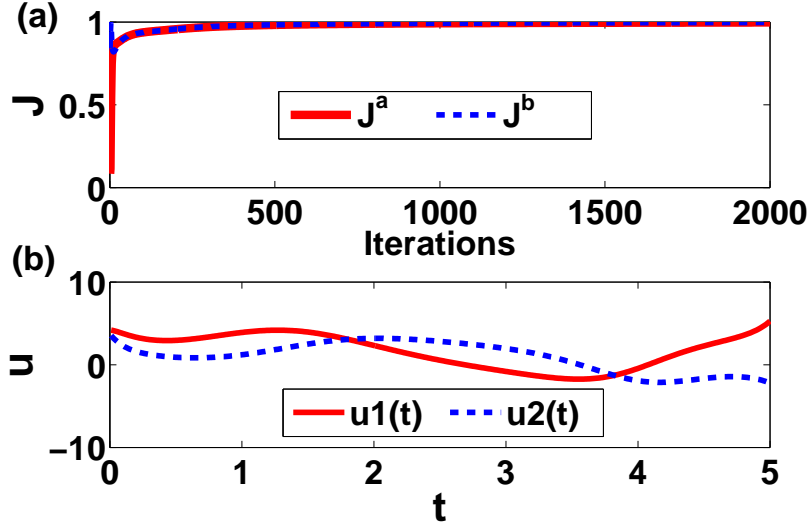}
\caption{Learning performance of discrimination between system
\emph{a} ($(\varepsilon^a_0,\varepsilon^a_u)=(0.9,0.9)$) and
system \emph{b} ($(\varepsilon^b_0,\varepsilon^b_u)=(1.1,1.1)$).
(a) Evolution of performance functions $J^a(u)$ and $J^b(u)$; (b)
The learned optimal control strategy $u(t)$.}
\end{figure}

\begin{figure}
\centering
\includegraphics[width=4.5in]{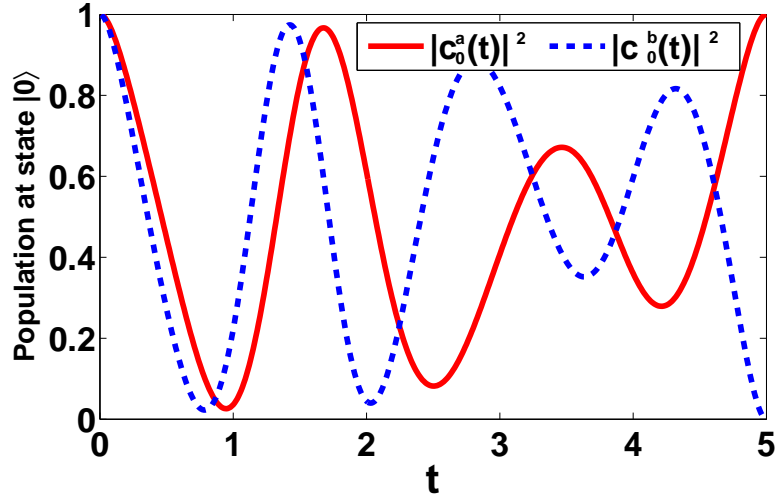}
\caption{Evolution of the states of system \emph{a}
($(\varepsilon^a_0,\varepsilon^a_u)=(0.9,0.9)$) and system
\emph{b} ($(\varepsilon^b_0,\varepsilon^b_u)=(1.1,1.1)$) regarding their
populations ($|c^a_0(t)|^2$ and $|c^b_0(t)|^2$) at the state
$|0\rangle$, respectively.}
\end{figure}

\begin{figure}
\centering
\includegraphics[width=4.5in]{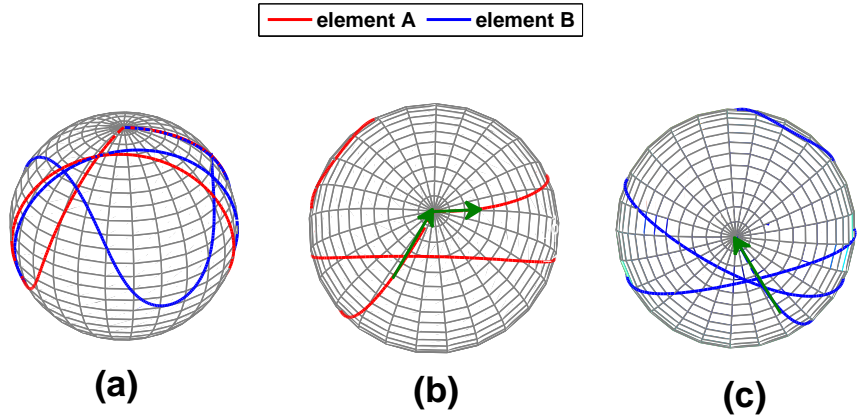}
\caption{Demonstration of the state transition trajectories of
system \emph{a} ($(\varepsilon^a_0,\varepsilon^a_u)=(0.9,0.9)$)
and system \emph{b}
($(\varepsilon^b_0,\varepsilon^b_u)=(1.1,1.1)$) on the Bloch sphere.}
\end{figure}

In the first example, two similar systems \emph{a} and \emph{b}
are characterized with parameters
$(\varepsilon^a_0,\varepsilon^a_u)=(0.9,0.9)$ and
$(\varepsilon^b_0,\varepsilon^b_u)=(1.1,1.1)$, respectively. The
numerical results are shown in Figs. 3-5. As shown in Fig. 3(a),
the learning process converges very quickly and the performance
function $J(u)$ converges to $0.999$ after about $2000$ steps of
iterative learning with an optimized control strategy
$u(t)=\{u_1(t), u_2(t)\}$ in Fig. 3(b). Then we apply the learned
optimal control strategy to systems \emph{a} and \emph{b}. The
evolution of their states can be clearly demonstrated regarding
their populations at the state $|0\rangle$ as shown in Fig. 4. At
time $t=T=5$, $|c^a_0(T)|^2=1.0000$ and $|c^b_0(T)|^2=0.0000$,
which indicates that, after the coherent control step, we can
discriminate system \emph{a} from system \emph{b} using a
projective measurement and the success probability is almost
$100\%$.

A 3-D visualized demonstration is further shown in Fig. 5 for the
state transition trajectories of systems \emph{a} and \emph{b} on the
Bloch sphere. For a two-level system, its
state can also be represented using a Bloch vector
$\mathbf{r}=(x,y,z)$ where $x=\text{tr}\{|\psi\rangle
\langle\psi|\sigma_{x}\}$, $y=\text{tr}\{|\psi\rangle
\langle\psi|\sigma_{y}\}$, $z=\text{tr}\{|\psi\rangle
\langle\psi|\sigma_{z}\}$ and $\text{tr}(\cdot)$ is the trace operator.
As shown in Fig. 5, using the same learned control strategy, the
state trajectories of systems \emph{a} and \emph{b} are
successfully driven from the same initial state
$|\psi_{0}\rangle=|0\rangle$ (i.e., $r_{0}=(0,0,1)$) to different
target states of $|\psi_\textrm{targetA}\rangle=|0\rangle$ (i.e.,
$r_\textrm{targetA}=(0,0,1)$) and
$|\psi_\textrm{targetB}\rangle=|1\rangle$ (i.e.,
$r_\textrm{targetA}=(0,0,-1)$), respectively.

\begin{figure}
\centering
\includegraphics[width=4.5in]{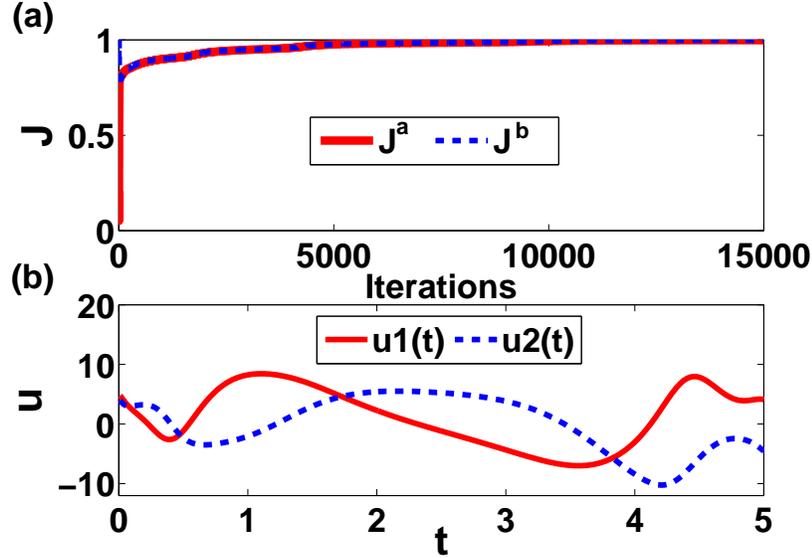}
\caption{Learning performance of discrimination between system
\emph{a} ($(\varepsilon^a_0,\varepsilon^a_u)=(0.95,0.95)$) and
system \emph{b}
($(\varepsilon^b_0,\varepsilon^b_u)=(1.05,1.05)$). (a) Evolution
of performance functions $J^a(u)$ and $J^b(u)$; (b) The learned
optimal control strategy $u(t)$.}
\end{figure}

\begin{figure}
\centering
\includegraphics[width=4.5in]{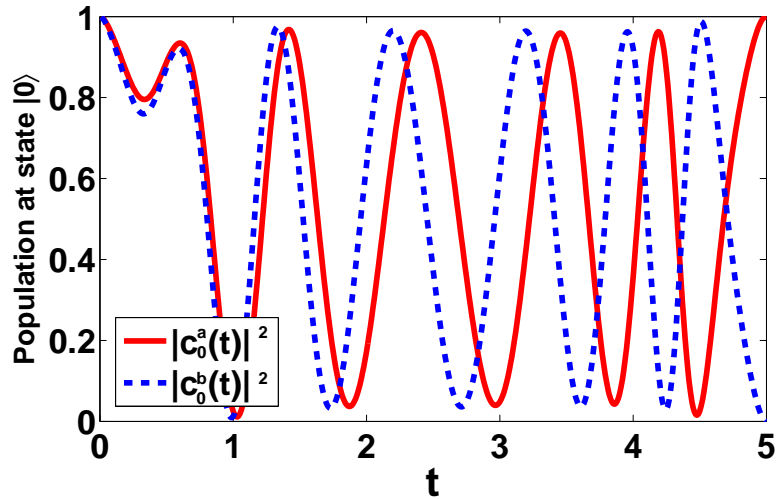}
\caption{Evolution of the states of system \emph{a}
($(\varepsilon^a_0,\varepsilon^a_u)=(0.95,0.95)$) and system
\emph{b} ($(\varepsilon^b_0,\varepsilon^b_u)=(1.05,1.05)$) regarding the
population at the state $|0\rangle$, respectively.}
\end{figure}

%\begin{figure}
%\centering
%\includegraphics[width=3.5in]{f11.eps}
%\caption{Demonstration of the state transition trajectories of
%element \emph{a} ($(\varepsilon^a_0,\varepsilon^a_u)=(0.95,0.95)$)
%and element \emph{b}
%($(\varepsilon^b_0,\varepsilon^b_u)=(1.05,1.05)$) on Bloch sphere.}
%\end{figure}

In the second example, two similar systems \emph{a} and \emph{b} are
characterized with parameters
$(\varepsilon^a_0,\varepsilon^a_u)=(0.95,0.95)$ and
$(\varepsilon^b_0,\varepsilon^b_u)=(1.05,1.05)$. The numerical
results are shown in Fig. 6 and Fig. 7. Similar to the first example, we can
also successfully learn an optimal control strategy (Fig. 6(b)) to
drive systems \emph{a} and \emph{b} to different target eigenstates
from the same initial state. The evolution of their states is shown
in Fig. 7 regarding their populations at the state $|0\rangle$,
respectively. By comparing the second example with the first one, it is clear
that the difference lies in the similarity between Hamiltonians. For the
second example, the increasing of the similarity of Hamiltonians makes
it more difficult to discriminate system \emph{a} from system \emph{b}.
More learning steps (about $15000$ steps) are needed for the second
example than the first one (around $2000$ steps). A larger control
strength is also needed for the second example than that in the first one (i.e., the amplitudes of controls in Fig. 6(b) are larger than
those in Fig. 3(b)). This phenomenon is comprehensively tested
through further numerical experiments with varied parameters. All of
the results show that the gradient flow based iterative learning
method is successful for discrimination of similar quantum systems
and also support the previous findings that optimal dynamic discrimination is feasible for many similar quantum systems in physics and chemistry
communities \cite{Beltrani et al 2009}, \cite{Li et al
2002}-\cite{Roslund et al 2011}.

\section{Quantum ensemble classification via SLC}\label{Sec4}
Binary classification is to classify the
members of a given set of objects into two classes on the basis of
whether they have certain properties or not. As introduced in Definition
\ref{BQEC}, for a binary QEC problem, we have to learn from a training
set as defined in Definition \ref{TrainingSet} and find out an
optimal control strategy for all the members in the quantum
ensemble. In this section, we combine a sampling-based learning control (SLC) approach into the quantum discrimination method introduced
above to solve the QEC problem (i.e., the
maximization problem formulated as Equation (\ref{ensemble
control})).

\subsection{SLC for quantum ensemble classification}\label{Subsec4.1}
The first key issue for QEC is how to obtain the training set.
Generally there are two ways to construct a training set for QEC:
(i) the data is provided initially and the training set can be
constructed directly, but we do not know the distribution of
parameters that characterize the members belonging to different
classes; (ii) no initial training data is provided but we know
the distribution of parameters and we can choose samples using the distribution information. The first way is very common in
classical machine learning problems, while the second way is
more suitable for the classification of quantum systems. In the quantum
domain, it is difficult to obtain a specific description for a
single system in a quantum ensemble, while we can characterize an ensemble of
similar systems with a distribution of parameters (e.g., Gaussian
distribution, Boltzmann distribution and uniform distribution).
According to the distribution of parameters for a quantum
ensemble, we can choose sample members to
construct the training set for the learning control design. This
approach is referred to as sampling-based learning control (SLC), which
originated in \cite{Chen et al 2013}, \cite{Dong et al 2013}
as a general framework for optimal control design of inhomogeneous
quantum ensembles and robust control design of quantum systems with uncertainties.

In the SLC approach, a generalized system is constructed by sampling
members from the inhomogeneous ensemble. In this paper, we adopt
the key idea from SLC and solve the supervised quantum learning
problem of QEC via constructing a generalized system using the
training set.

Suppose we have obtained a training set $D_N=\{(H^n(t),y_n)\}$
($n=1,2,\ldots,N$) for the binary QEC problem, where $y_n \in \{A,
B\}$ and $H^n(t)$ is the time-dependent Hamiltonian that describes
the $n\text{th}$ member of the quantum ensemble. Now we split $D_N$
into two subsets according to the value of $y_n$ and rewrite the
training set as follows
\begin{equation}\label{VariedTS}
\begin{split}
& D_N=D_{N_A}\cup D_{N_B},\ N=N_A+N_B,\\
& D_{N_A}=\{(H^A_{\varepsilon^{n_A}_0,
\varepsilon^{n_A}_u}(t),y_{n_A}=A)\},\ n_A=1,2,\ldots,N_A,\\
& D_{N_B}=\{(H^B_{\varepsilon^{n_B}_0,
\varepsilon^{n_B}_u}(t),y_{n_B}=B)\},\ n_B=N_A+1,N_A+2,\ldots,N,\\
\end{split}
\end{equation}
where $$H^A_{\varepsilon^{n_A}_0,
\varepsilon^{n_A}_u}(t)=g^A_0(\varepsilon^{n_A}_0)H_{0}+g^A_u(\varepsilon^{n_A}_u)\sum_{m=1}^{M}u_{m}(t)H_{m}$$
and $$H^B_{\varepsilon^{n_B}_0,
\varepsilon^{n_B}_u}(t)=g^B_0(\varepsilon^{n_B}_0)H_{0}+g^B_u(\varepsilon^{n_B}_u)\sum_{m=1}^{M}u_{m}(t)H_{m}.$$

Using the training set (\ref{VariedTS}), we can construct a
generalized system as follows
\begin{equation}\label{generalized-system}\small
\frac{d}{dt}\left(%
\begin{array}{c}
  |{\psi}^A_{\varepsilon^1_0,
\varepsilon^1_u}(t)\rangle \\
\vdots \\
  |{\psi}^A_{\varepsilon^{N_A}_0,
\varepsilon^{N_A}_u}(t)\rangle \\
  |{\psi}^B_{\varepsilon^{N_A+1}_0,
\varepsilon^{N_A+1}_u}(t)\rangle \\
  \vdots \\
  |{\psi}^B_{\varepsilon^{N}_0,
\varepsilon^{N}_u}(t)\rangle \\
\end{array}%
\right)
=-i\left(%
\begin{array}{c}
  H^A_{\varepsilon^1_0,
\varepsilon^1_u}(t)|\psi^A_{\varepsilon^1_0,
\varepsilon^1_u}(t)\rangle \\
\vdots \\
  H^A_{\varepsilon^{N_A}_0,
\varepsilon^{N_A}_u}(t)|\psi^A_{\varepsilon^{N_A}_0,
\varepsilon^{N_A}_u}(t)\rangle \\
H^B_{\varepsilon^{N_A+1}_0,
\varepsilon^{N_A+1}_u}(t)|\psi^B_{\varepsilon^{N_A+1}_0,
\varepsilon^{N_A+1}_u}(t)\rangle \\
  \vdots \\
  H^B_{\varepsilon^{N}_0,
\varepsilon^{N}_u}(t)|\psi^B_{\varepsilon^{N}_0,
\varepsilon^{N}_u}(t)\rangle \\
\end{array}%
\right).
\end{equation}
The performance function for this generalized system is defined by
\begin{equation}\label{eq:cost}
J_N(u):=w_AJ^A+w_BJ^B,
\end{equation}
where
\begin{equation}\small \label{eq:cost1}
\begin{split}
&J^A=\frac{1}{N_A}\sum_{n_A=1}^{N_A} J^A_{\varepsilon^{n_A}_0,
\varepsilon^{n_A}_u}(u)=\frac{1}{N_A}\sum_{n_A=1}^{N_A}\vert
\langle\psi^A_{\varepsilon^{n_A}_0,
\varepsilon^{n_A}_u}(T)|\psi_{\text{targetA}}\rangle\vert^{2},\\
&J^B=\frac{1}{N_B}\sum_{n_B=N_A+1}^{N} J^B_{\varepsilon^{n_B}_0,
\varepsilon^{n_B}_u}(u)=\frac{1}{N_B}\sum_{n_B=N_A+1}^{N}\vert
\langle\psi^B_{\varepsilon^{n_B}_0,
\varepsilon^{n_B}_u}(T)|\psi_{\text{targetB}}\rangle\vert^{2}.\\
\end{split}
\end{equation}

The task of the training step is to find a control
strategy that maximizes the performance function defined in
\eqref{eq:cost}. From equations \eqref{gradientJ},
\eqref{eq:cost} and \eqref{eq:cost1}, we have
\begin{equation}\small \label{gradientJsemble}
\begin{split} \nabla J_N(u^{k}_m)= &
\frac{2w_a}{N_A}\sum_{n_A=1}^{N_A}\Im\left(\langle\psi^A_{\varepsilon^{n_A}_0,\varepsilon^{n_A}_u}(T)\vert\psi_{\textrm{targetA}}\rangle\langle\psi_{\textrm{targetA}}\vert
G^A_{n_A,m}(t)\vert\psi_0\rangle\right)\\
&
+\frac{2w_b}{N_B}\sum_{n_B=N_A+1}^{N}\Im\left(\langle\psi^B_{\varepsilon^{n_B}_0,\varepsilon^{n_B}_u}(T)\vert\psi_{\textrm{targetB}}\rangle\langle\psi_{\textrm{targetB}}\vert
G^B_{n_B,m}(t)\vert\psi_0\rangle\right)
\end{split}
\end{equation}
where
$$G^A_{n_A,m}(t)=U_{\varepsilon^{n_A}_0,\varepsilon^{n_A}_u}(T)U_{\varepsilon^{n_A}_0,\varepsilon^{n_A}_u}^\dagger(t)g^A_u(\varepsilon^{n_A}_u)H_mU_{\varepsilon^{n_A}_0,\varepsilon^{n_A}_u}(t),$$
$$G^B_{n_B,m}(t)=U_{\varepsilon^{n_B}_0,\varepsilon^{n_B}_u}(T)U_{\varepsilon^{n_B}_0,\varepsilon^{n_B}_u}^\dagger(t)g^B_u(\varepsilon^{n_B}_u)H_mU_{\varepsilon^{n_B}_0,\varepsilon^{n_B}_u}(t).$$
Then we design the SLC algorithm (\emph{Algorithm 2}) for binary QEC using the gradient flow
method to approximate an optimal control strategy
$u^{*}=\{u^{*}_{m}(t)\}$.

\begin{algorithm}
\caption{SLC for binary QEC} \label{SLCforQEC}
\begin{algorithmic}[1]

\State Set the index of iterations $k=0$ \State Choose a set of
arbitrary controls $u^{k=0}(t)=\{u_{m}^{0}(t),\ m=1,2,\ldots,M\}, t
\in [0,T]$

\Repeat {\ (for each iterative process)}

\Repeat {\ (for each member in training subset $D_{N_A}$,
$n_A=1,2,\ldots,N_A$ )} \State Compute the propagator
$U_{\varepsilon^{n_A}_0,\varepsilon^{n_A}_u}^{k}(t)$ with the
control strategy $u^{k}(t)$ \Until {\ $n_A=N_A$}

\Repeat {\ (for each member in training subset $D_{N_B}$,
$n_B=N_A+1,N_A+2,\ldots,N$ )} \State Compute the propagator
$U_{\varepsilon^{n_B}_0,\varepsilon^{n_B}_u}^{k}(t)$ with the
control strategy $u^{k}(t)$ \Until {\ $n_B=N$}

\Repeat {\ (for each control $u_{m}(t)$ ($m=1,2,\ldots,M$) of the
control vector $u^k(t)$)}

\State $\delta_m^{k}(t):=\nabla J_N(u^{k}_m)$ and compute $\nabla
J_N(u^{k}_m)$ using equation \eqref{gradientJsemble}

\State $u_{m}^{k+1}(t)=u_{m}^{k}(t)+\eta_{k} \delta_{m}^{k}(t)$

\Until {\ $m=M$}

\State $k=k+1$

\Until {\ the learning process ends}

\State The optimal control strategy
$u^{*}(t)=\{u_{m}^*(t)\}=\{u_{m}^{k}(t)\}, \ m=1,2,\ldots,M$

\end{algorithmic}
\end{algorithm}

\begin{remark}
Both \emph{Algorithm 1} and \emph{Algorithm 2} are developed using
the gradient flow method. Their convergence is closely related to
quantum control problems. According to the theory of quantum control
landscape \cite{Rabitz et al 2004}, \cite{Chakrabarti and Rabitz
2007}, gradient-based algorithms are effective for trap-free quantum
optimal control problems and many practical quantum control problems
are trap-free problems \cite{Jirari and Potz 2005}-\cite{Roslund and
Rabitz 2009}. The classification considered in this
paper is trap-free. For those complex quantum control problems
(e.g., control of multi-level open quantum systems),
stochastic learning techniques (e.g., genetic algorithms) may be
required.
\end{remark}

\begin{remark}
As for the specific techniques of choosing samples ($N$ members
of the ensemble), we generally choose them according to the
functions $g^A_0(\cdot)$, $g^A_u(\cdot)$,
$g^B_0(\cdot)$ and $g^B_u(\cdot)$, and the
distribution of the inhomogeneity parameters $\varepsilon_0$ and
$\varepsilon_u$. It is clear that the basic motivation of the
proposed sampling-based learning control approach is to design a control law
using only a few sampling members instead of the whole ensemble
(consisting of a large number of members). Therefore, it is
necessary to choose the samples that are representative for
the quantum ensemble. Generally if we know the distribution
of the parameter dispersion, it is practical and convenient to
choose artificial members for the construction of the generalized
system. In numerical examples, we will
demonstrate the detailed method for choosing samples and more
related topics have also been discussed in \cite{Chen et al 2013},
\cite{Dong et al 2013}. In addition, overlapped distributions of the
inhomogeneity parameters for class \emph{A} and class \emph{B} may
lead to the problem of overlapped classification, which is a
challenging task even for classical classification problems
\cite{Liu 2009}. In the next subsection, we demonstrate the
classification performance for both cases with and without
class overlapping.
\end{remark}

\subsection{Numerical examples}\label{Subsec4.2}
We consider two-level quantum systems. For two similar classes of members in an inhomogeneous
quantum ensemble, the Hamiltonians can be described as
\begin{equation}
\begin{split}
&H^A_{\varepsilon_0,\varepsilon_u}(t)=\frac{1}{2}g^A_0(\varepsilon_0)\sigma_{z}+\frac{1}{2}g^A_u(\varepsilon_u)(u_{1}(t)\sigma_{x}+u_{2}(t)\sigma_{y}),\\
&H^B_{\varepsilon_0,\varepsilon_u}(t)=\frac{1}{2}g^B_0(\varepsilon_0)\sigma_{z}+\frac{1}{2}g^B_u(\varepsilon_u)(u_{1}(t)\sigma_{x}+u_{2}(t)\sigma_{y}).\\
\end{split}
\end{equation}
Assume $g^A_0(\varepsilon_0)=\varepsilon_0$ with distribution
$d^A_0(\varepsilon_0)$, $g^A_u(\varepsilon_u)=\varepsilon_u$ with
distribution $d^A_u(\varepsilon_u)$,
$g^B_0(\varepsilon_0)=\varepsilon_0$ with distribution
$d^B_0(\varepsilon_0)$, and $g^B_u(\varepsilon_u)=\varepsilon_u$ with
distribution $d^B_u(\varepsilon_u)$.

Suppose the distributions of $\varepsilon_0$ and $\varepsilon_u$ for
class \emph{A} are
$d^A_0(\varepsilon_0)=\Phi(\frac{\varepsilon_0-\mu^A_0}{\sigma^A_0})$
and
$d^A_u(\varepsilon_u)=\Phi(\frac{\varepsilon_u-\mu^A_u}{\sigma^A_u})$,
respectively, where $\Phi(x)=\int^x_{-\infty}\frac{1}{\sqrt{2\pi}}\text{exp}(-\frac{1}{2}\nu^2)d\nu$ is the
distribution function of the standard normal distribution. We
may choose some equally spaced samples in the
$\varepsilon_0-\varepsilon_u$ space. For example, we may choose the
intervals of $[\mu^A_0-3\sigma^A_0, \mu^A_0+3\sigma^A_0]$ and
$[\mu^A_u-3\sigma^A_u, \mu^A_u+3\sigma^A_u]$, and divide them into
$N^A_{\varepsilon_0}+1$ and $N^A_{\varepsilon_u}+1$ subintervals,
respectively, where $N^A_{\varepsilon_0}$ and $N^A_{\varepsilon_u}$
are usually positive odd numbers. Then the number of samples for
class \emph{A} is $N_A=N^A_{\varepsilon_0}N^A_{\varepsilon_u}$,
where $\varepsilon^{n_A}_0$ and $\varepsilon^{n_A}_u$ can be chosen
from the combination of $(\varepsilon_{0}^{n^0_{A}},
\varepsilon_{u}^{n^u_{A}})$ as follows

\begin{equation}\label{discreteA}
\left\{ \begin{array}{c} \varepsilon^{n_A}_0 \in
\{\varepsilon_{0}^{n^0_{A}}=\mu^A_0-3\sigma^A_0+\frac{(2n^0_{A}-1)3\sigma^A_0}{N^A_{\varepsilon_0}},
\ n^0_{A}=1,2,\ldots, N^A_{\varepsilon_0}\},\\
\varepsilon^{n_A}_u \in
\{\varepsilon_{u}^{n^u_{A}}=\mu^A_u-3\sigma^A_u+\frac{(2n^u_{A}-1)3\sigma^A_u}{N^A_{\varepsilon_u}},\
\
n^u_{A}=1,2,\ldots, N^A_{\varepsilon_u}\}. \\
\end{array}
\right.
\end{equation}
In practical applications, the numbers of $N^A_{\varepsilon_0}$
and $N^A_{\varepsilon_u}$ can be chosen by experience or be tried
through numerical computation. As long as the generalized system
can model the quantum ensemble and is effective to find the
optimal control strategy, we prefer smaller numbers
$N^A_{\varepsilon_0}$ and $N^A_{\varepsilon_u}$ to speed up the
training process and simplify the generalized system. A similar
expression to \eqref{discreteA} defines the samples for class
\emph{B}. We use the performance function as defined in
\eqref{eq:cost} with $w_A=w_B=0.5$. Now we use \emph{Algorithm 2}
to find the optimal control strategy.
%For a given threshold
%$\epsilon>0$, if $|J_N(u^{k+1})-J_N(u^{k})|<\epsilon$ for
%uninterrupted $n_e$ learning steps, we may think we find a
%suitable control strategy for the generalized system for QEC. In
%this paper, we set $\epsilon=10^{-4}$ and $n_e=100$ for all the
%numerical experiments.

The parameter settings are listed as follows: $w_A=w_B=0.5$, the
initial state for each member of the quantum ensemble
$C_{0}=(1,0)^{T}$, i.e., $|\psi_{0}\rangle=|0\rangle$, and the
target state for members belonging to class \emph{A}
$C_{\text{targetA}}=(1,0)^{T}$, i.e.,
$|\psi_{\text{targetA}}\rangle=|0\rangle$; the target state for
elements belonging to class \emph{B}
$C_{\text{targetB}}=(0,1)^{T}$, i.e.,
$|\psi_{\text{targetB}}\rangle=|1\rangle$; The ending time $T=8$
(in atomic units) and the total time duration $[0,T]$ is equally
discretized into $Q=800$ time slices with each time slice $\Delta
t=(t_{q}-t_{q-1})|_{q=1,2,\ldots,Q}=T/Q=0.01$;
$N^A_{\varepsilon_0}=N^A_{\varepsilon_u}=N^B_{\varepsilon_0}=N^B_{\varepsilon_u}=5$;
the learning rate $\eta_{k}=0.2$; the control strategy is
initialized as $u^{k=0}(t)=\{u^{0}_{1}(t)=\sin t,
u^{0}_{2}(t)=\sin t\}$.

In the training step, we use $J(u)$ as the performance function
which represents the measure of weighted accuracy
for QEC. After we apply the optimized control
$u^{*}$ to the inhomogeneous quantum ensemble, we use fidelity to characterize how well
every member is classified. The fidelity
between the final state $|\psi^A_{\varepsilon_{0},\varepsilon_{u}}(T)\rangle$ of a member belonging to
class \emph{A} and the
target state $|\psi_{\text{targetA}}\rangle$ is defined as follows
\cite{Nielsen and Chuang 2000}
\begin{equation}\label{fidelity}
F(|\psi^A_{\varepsilon_{0},\varepsilon_{u}}(T)\rangle,|\psi_{\text{targetA}}\rangle)=|\langle
\psi^A_{\varepsilon_{0},\varepsilon_{u}}(T)|\psi_{\text{targetA}}\rangle|.
\end{equation}
A similar representation can be defined for the final state $|\psi^B_{\varepsilon_{0},\varepsilon_{u}}(T)\rangle$
of a member belonging to class \emph{B}
 and the
target state $|\psi_{\text{targetB}}\rangle$. It is clear that the
accuracy of QEC can be calculated with
\begin{equation}
\begin{array}{cc}
\zeta&=J(u)=\frac{1}{2}(\mathbb{E}[J^A]+\mathbb{E}[J^B])\ \ \ \ \ \ \ \ \\
&=\frac{1}{2}(\mathbb{E}[F^2(|\psi^A_{\varepsilon_{0},\varepsilon_{u}}(T)\rangle,|\psi_{\text{targetA}}\rangle)]\\
&+\mathbb{E}[F^2(|\psi^B_{\varepsilon_{0},\varepsilon_{u}}(T)\rangle,|\psi_{\text{targetB}}\rangle)]).\\
\end{array}
\end{equation}

We demonstrate and analyze several groups of numerical examples
for the cases (1), (2) and (3) with class overlapping, whose
parameters characterizing the inhomogeneity are listed as in Table
I.
%\uppercase\expandafter{\romannumeral1}.

\begin{table}[!htb]
\scalebox{0.85}{
\begin{tabular}{|c|c|c|c|c|}
\hline
Distribution & $d^A_0(\varepsilon_0)=$ & $d^A_u(\varepsilon_u)=$ & $d^B_0(\varepsilon_0)$ & $d^B_u(\varepsilon_u)=$\\
 Function & $\Phi(\frac{\varepsilon_0-\mu^A_0}{\sigma^A_0})$ & $\Phi(\frac{\varepsilon_u-\mu^A_u}{\sigma^A_u})$ & $\Phi(\frac{\varepsilon_0-\mu^B_0}{\sigma^B_0})$ & $=\Phi(\frac{\varepsilon_u-\mu^B_u}{\sigma^B_u})$\\
 & & & & \\
\hline
 & ($\mu^A_0,3\sigma^A_0$) & ($\mu^A_u,3\sigma^A_u$) & ($\mu^B_0,3\sigma^B_0$) & ($\mu^B_u,3\sigma^B_u$)\\
\hline
case (1) & (0.85,0.05) & (0.85,0.05) & (1.15,0.05) & (1.15,0.05)\\
\hline
case (2) & (0.85,0.15) & (0.85,0.15) & (1.15,0.15) & (1.15,0.15)\\
\hline
case (3) & (0.80,0.05) & (0.80,0.05) & (1.20,0.05) & (1.20,0.05)\\
\hline
\end{tabular}}
\caption{The parameters characterizing the inhomogeneity distribution.}
\end{table}

The learning control performance for \emph{case (1)} is shown in
Fig. 8 and Fig. 9. As shown in Fig. 8, the learning algorithm
converges quickly after about $8000$ steps of iterations and finds an
optimized control for the coherent control step of binary QEC.
Applying the learned control to $300$ randomly selected testing
samples ($150$ for \emph{class A} and $150$ for \emph{class B}), the
control performance is shown in Fig. 9.
The mean value of fidelity for the testing of \emph{class A} is $0.9976$
and for \emph{class B} is $0.9985$. With additional $10^4$ testing
samples for both \emph{class A} and \emph{class B}, the
classification accuracy in \emph{case (1)} is estimated as
$\zeta=99.62\%$.
% (the limit of the best classification accuracy is almost $1$).

\begin{figure}
\centering
\includegraphics[width=4.5in]{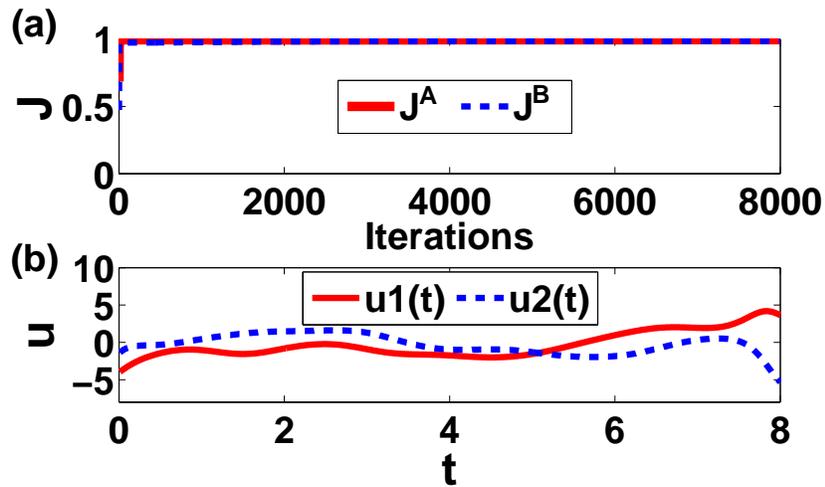}
\caption{Learning performance of binary QEC for \emph{case (1)}: (a)
evolution of performance function $J^A$ and $J^B$; (b) the learned
optimal control for QEC.}
\end{figure}

\begin{figure}
\centering
\includegraphics[width=4.5in]{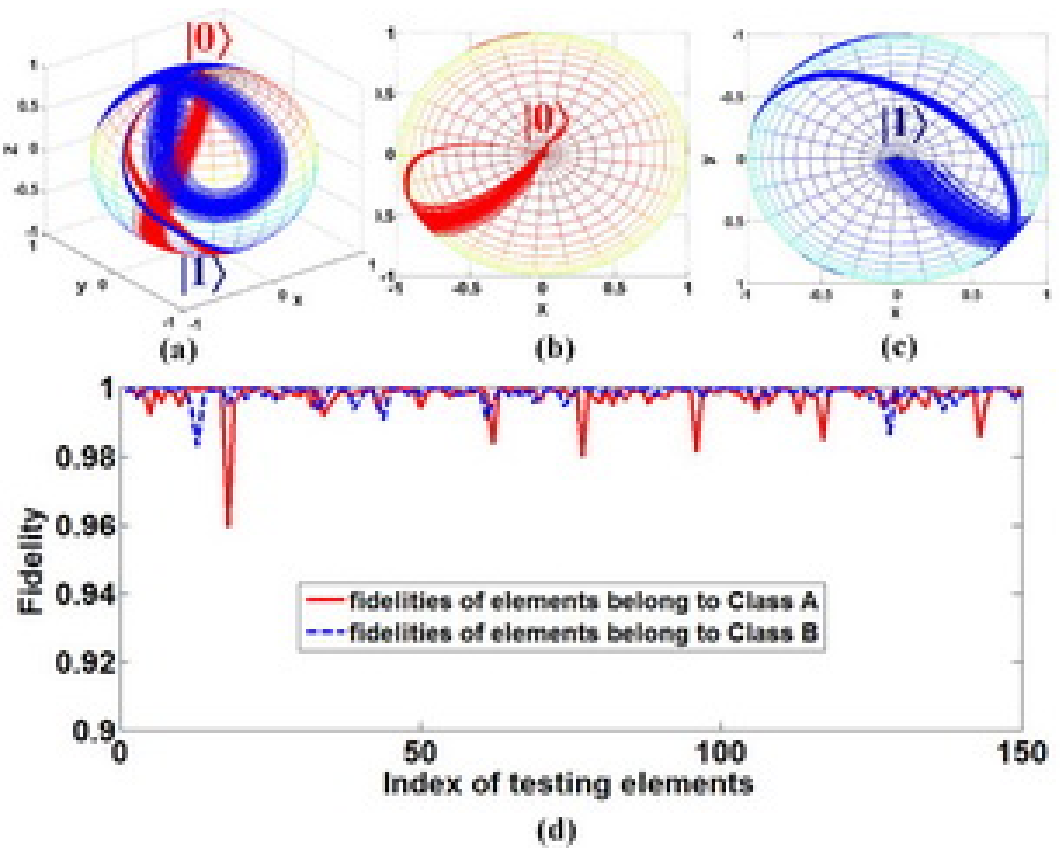}
\caption{Control performance of binary QEC for \emph{case (1)}: (a)
demonstration of the state transition of all members on the Bloch
sphere using the same learned control; (b) trajectories of state
transition for members in \emph{class A} from
$|\psi_{0}\rangle=|0\rangle$ to
$|\psi_{\text{targetA}}\rangle=|0\rangle$; (c) trajectories of state
transition for members in \emph{class B} from
$|\psi_{0}\rangle=|0\rangle$ to
$|\psi_{\text{targetB}}\rangle=|1\rangle$; (d) control performance
regarding fidelity.}
\end{figure}

\begin{figure}
\centering
\includegraphics[width=4.5in]{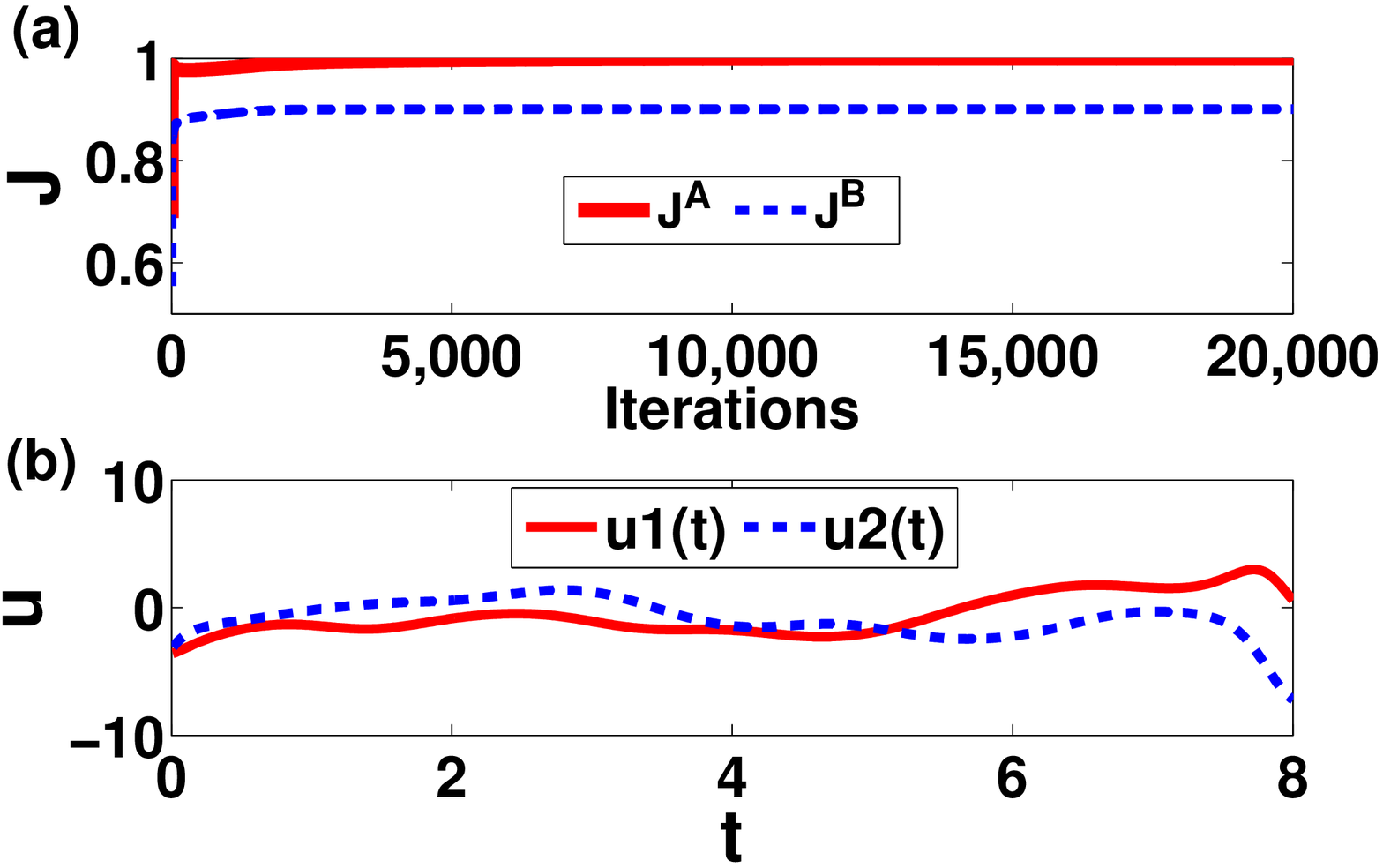}
\caption{Learning performance of binary QEC in \emph{case (2)}:
(a) evolution of performance function $J^A$ and $J^B$; (b) the
learned optimal control for QEC.}
\end{figure}

\begin{figure}
\centering
\includegraphics[width=4.5in]{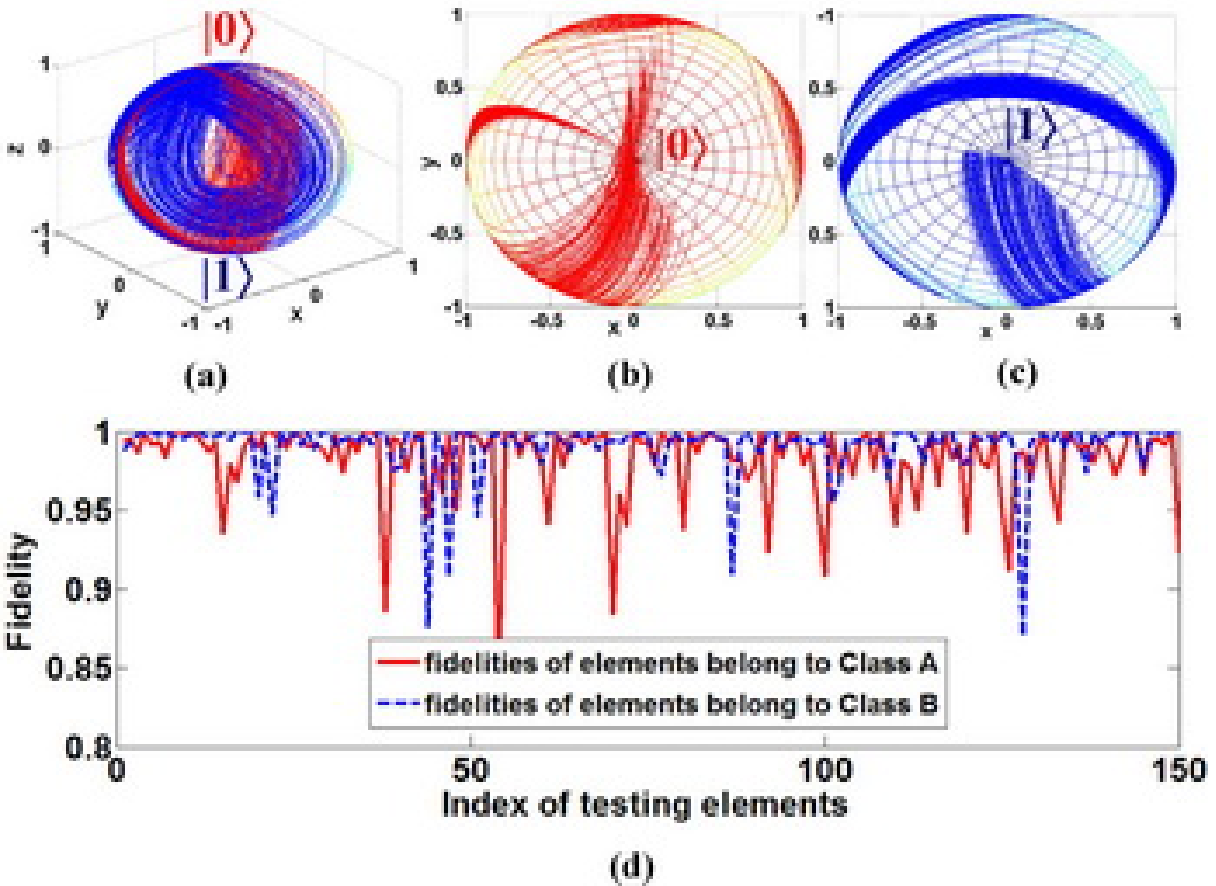}
\caption{Control performance of binary QEC for \emph{case (2)}:
(a) demonstration of the state transition of all members on the Bloch
sphere using the same learned control; (b) trajectories of state
transition for members in \emph{class A} from
$|\psi_{0}\rangle=|0\rangle$ to
$|\psi_{\text{targetA}}\rangle=|0\rangle$; (c) trajectories of
state transition for members in \emph{class B} from
$|\psi_{0}\rangle=|0\rangle$ to
$|\psi_{\text{targetB}}\rangle=|1\rangle$; (d) control performance
regarding fidelity.}
\end{figure}

Compared with \emph{case (1)}, we study the effect of larger
dispersion on the Hamiltonian in \emph{case (2)} with larger deviation.
The learning control performance for \emph{case (2)} is shown in
Fig. 10 and Fig. 11. As shown in Fig. 10(a), many more learning
steps are needed to find a satisfactory control and the learned
optimal control is shown in Fig. 10(b). Due to a larger dispersion,
the control performance is a little worse than that in \emph{case
(1)}. The mean value of fidelity for the testing of
\emph{class A} is $0.9821$ and for \emph{class B} is $0.9905$. With an
additional $10^4$ testing samples for both \emph{class A} and
\emph{class B}, the classification accuracy in \emph{case (2)} is
estimated as $\zeta=97.35\%$.
% (the limit of the best classification
%accuracy is almost $99.865\%$).

\begin{figure}
\centering
\includegraphics[width=4.5in]{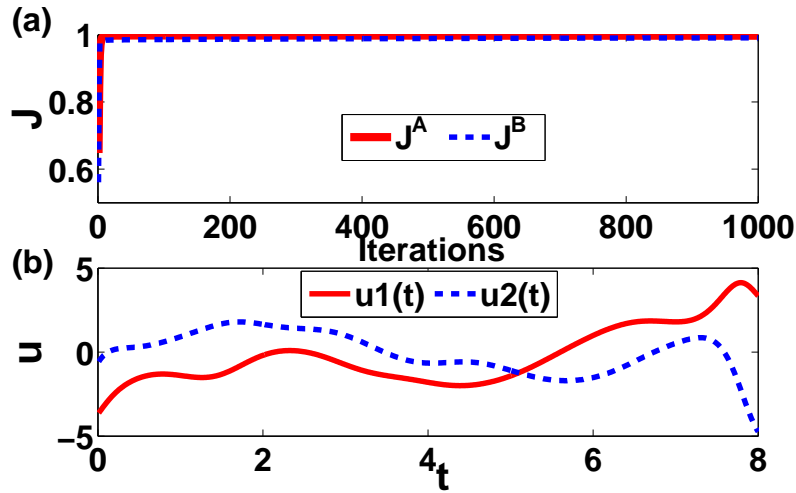}
\caption{Learning performance of binary QEC in \emph{case (3)}:
(a) evolution of performance function $J^A$ and $J^B$; (b) the
learned optimal control for QEC.}
\end{figure}

\begin{figure}
\centering
\includegraphics[width=4.5in]{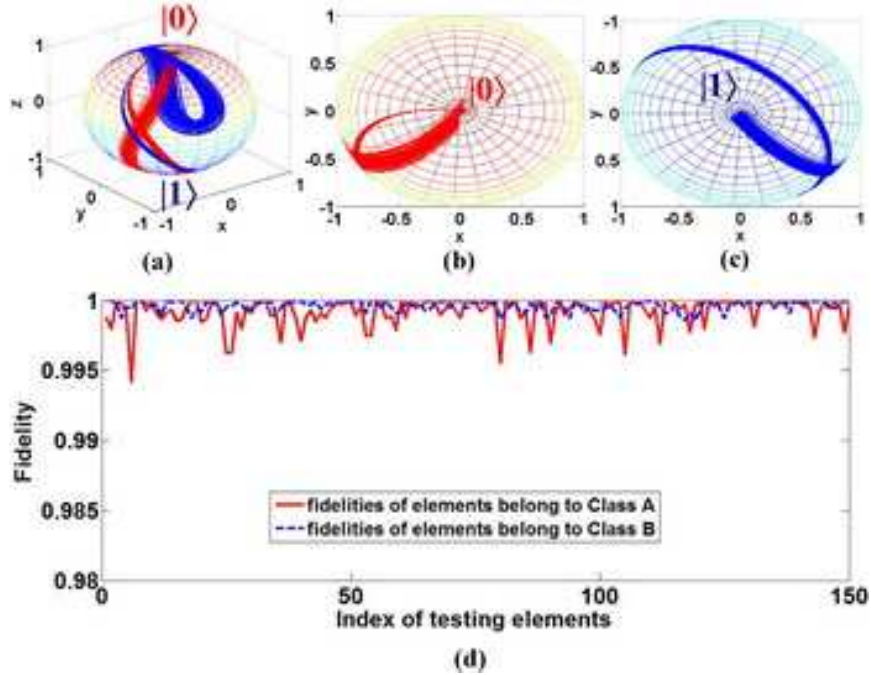}
\caption{Control performance of binary QEC in \emph{case (3)}:
(a) demonstration of the state transition of all members on the Bloch
sphere using the same learned control; (b) trajectories of state
transition for members in \emph{class A} from
$|\psi_{0}\rangle=|0\rangle$ to
$|\psi_{\text{targetA}}\rangle=|0\rangle$; (c) trajectories of
state transition for members in \emph{class B} from
$|\psi_{0}\rangle=|0\rangle$ to
$|\psi_{\text{targetB}}\rangle=|1\rangle$; (d) control performance
regarding fidelity.}
\end{figure}

In \emph{case (3)}, we further study the effect of a larger difference
of Hamiltonian between class \emph{A} and class \emph{B}. The
difference between the means of the distribution in \emph{case (1)}
$|\mu_0^A-\mu_0^B|=0.3$ and $|\mu_u^A-\mu_u^B|=0.3$, while in
\emph{case (3)} $|\mu_0^A-\mu_0^B|=0.4$ and $|\mu_u^A-\mu_u^B|=0.4$.
The learning control performance for \emph{case (3)} is shown in
Fig. 12 and Fig. 13. As shown in Fig. 12(a), much fewer learning
steps are needed to find a satisfactory control and the learned
optimal control is shown in Fig. 12(b). Due to larger difference
between class \emph{A} and class \emph{B}, the control performance
is better than that in \emph{case (1)}. The mean
value of fidelity for testing of \emph{class A} is $0.9992$ and for
\emph{class B} is $0.9996$. With additional $10^4$ testing samples
for both \emph{class A} and \emph{class B}, the classification
accuracy in \emph{case (3)} is estimated as $\zeta=99.88\%$.
% (we can calculate using the parameter distributions that the limit of the
%best classification accuracy is almost $1$).

From the numerical results in the above three cases, we have the following conclusions: first, the SLC approach is effective for the
binary QEC problem and can achieve a high level of classification accuracy;
second, the classification performance is deteriorated with
larger dispersion on the Hamiltonian and smaller difference of
Hamiltonians between class \emph{A} and class \emph{B}. More numerical experiments are carried out and similar findings
are obtained. For ease of demonstration,
we may use the same deviation $\sigma$ for all the distributions in a certain case. Define
the dispersion on the Hamiltonian as $\text{Disp}=3\sigma$ and define
the difference of the Hamiltonian between class \emph{A} and class
\emph{B} as
$\text{Diff}=\frac{1}{2}(|\mu^A_0-\mu^B_0|+|\mu^A_u-\mu^B_u|)$. The
collective results are shown in Fig. 14 with a 3D Pareto front. For a
detailed discussion about Pareto front, please refer to \cite{Beltrani et al
2009}.

\begin{figure}
\centering
\includegraphics[width=4.5in]{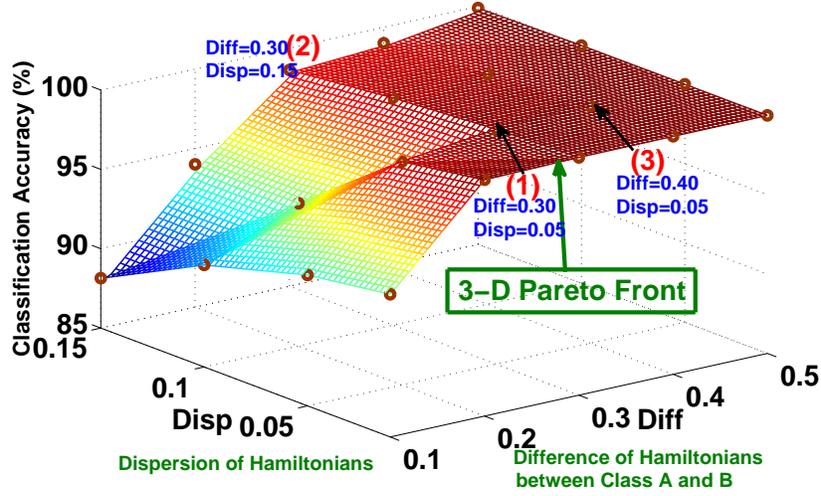}
\caption{Three-dimensional Pareto front of binary QEC.}
\end{figure}

\begin{remark}
In this paper, the classification problem under consideration involves class overlapping, which is more challenging than that
without class overlapping. The proposed approach can be easily
applied to the QEC problem without class overlapping and can obtain
even better performance. For example, in \emph{case (1)}
its counterpart without class overlapping
can be characterized with truncated normal distribution. Let the
probability density function of a truncated normal distribution be
$$p(x,\mu,\sigma,l,r)=\frac{\frac{1}{\sigma}\phi(\frac{x-\mu}{\sigma})}{\Phi(\frac{r-\mu}{\sigma})
-\Phi(\frac{l-\mu}{\sigma})}$$
where $\phi(x)$ is the probability density function of the standard normal distribution. The probability density functions for the truncated normal
distributions are set as follows:
$$p_0^A=p(\varepsilon_0,\mu^A_0,\sigma^A_0,-\infty,\overline{\mu_0}),\
p_u^A=p(\varepsilon_u,\mu^A_u,\sigma^A_u,-\infty,\overline{\mu_u}),$$
$$p_0^B=p(\varepsilon_0,\mu^B_0,\sigma^B_0,\overline{\mu_0},+\infty),\ p_u^B=p(\varepsilon_u,\mu^B_u,\sigma^B_u,\overline{\mu_u},+\infty).$$
where $\overline{\mu_0}=\frac{\mu_0^A+\mu_0^B}{2}$ and
$\overline{\mu_u}=\frac{\mu_u^A+\mu_u^B}{2}$. Using the same
approach and parameter settings in \emph{case (1)}, we can achieve the
classification accuracy $\zeta'=99.66\%$. Similar comparison is investigated in \emph{case (2)} and \emph{case (3)}.
These results are presented in Table II.
%(compared with
%$\zeta=99.62\%$ for  with class overlapping).
%Similarly we can further get $\zeta'=97.70\%$ for the case without
%class overlapping for \emph{case (2)} (with class overlapping
%$\zeta=97.35\%$) and $\zeta'=99.92\%$ for \emph{case (3)} (with
%class overlapping $\zeta=99.88\%$).
\end{remark}

\begin{table}[!htb]
\scalebox{0.85}{
\begin{tabular}{|c|c|c|}
\hline
case & accuracy $\zeta$ & accuracy $\zeta'$\\
& with class overlapping & without class overlapping\\
\hline
case (1) & $\zeta=99.62\%$ & $\zeta'=99.66\%$ \\
\hline
case (2) & $\zeta=97.35\%$ & $\zeta'=97.70\%$ \\
\hline
case (3) & $\zeta=99.88\%$ & $\zeta'=99.92\%$ \\
\hline
\end{tabular}}
\caption{A comparison between the classification accuracy $\zeta$ with class overlapping
and the classification accuracy $\zeta'$ without class overlapping for different cases.}
\end{table}

\section{Multiclass classification of multi-level quantum ensembles}\label{Sec5}
In machine learning, multiclass classification involves classifying instances into more than two classes.
Some classification algorithms naturally permit the use of
more than two classes \cite{Lin et al 2013TNNLS}. A useful strategy is the one-vs-all strategy,
where a single classifier is trained per class to distinguish that
class from all other classes \cite{Rifkin and Klautau 2004}.

The SLC based QEC approach proposed for QEC in Section \ref{Sec4} can be extended to multiclass classification of multilevel quantum ensembles using
the one-vs-all strategy. For example, an inhomogeneous quantum
ensemble consists of three classes of members (i.e., classes
\emph{A}, \emph{B} and \emph{C}). First, by applying the binary QEC
approach introduced in Section \ref{Sec4}, we can classify them into
two classes (one for members belonging to class \emph{A} and the
other for all the members belonging to classes \emph{B} and \emph{C}). Then we use the
binary QEC approach again to classify the members belonging to
class \emph{B} from the members belonging to class \emph{C}.
According to the numerical results demonstrated in Section
\ref{Sec4}, good classification performance is also expected for
these multiclass classification problems. However,
the one-vs-all strategy may cause additional cost since the process involves
multiple times of binary classification and multiple learning control procedures.

In this section we use a different strategy from the one-vs-all strategy to extend the proposed SLC based classification
method to multiclass classification of multi-level quantum ensembles. In this strategy, only one time of quantum coherent
control procedure is needed to
implement the multiclass QEC.

We consider an inhomogeneous quantum ensemble of three-level
$\Lambda$-type atomic systems \cite{Chen et al 2013}. The evolving state $|\psi(t)\rangle$
of the $\Lambda$-type system can be expanded in terms of the
eigenstates as follows:
\begin{equation}\label{level3system1}
|\psi(t)\rangle=
c_{1}(t)|1\rangle+c_{2}(t)|2\rangle+c_{3}(t)|3\rangle,
\end{equation}
where $|1\rangle$, $|2\rangle$ and $|3\rangle$ are the basis states
of the lower, middle and upper atomic states, respectively,
corresponding to the free Hamiltonian
\begin{equation}\small \label{level3system2}
H_{0}=
\left(%
\begin{array}{ccc}
  1.5 & 0 & 0 \\
  0 & 1  & 0 \\
  0 & 0  & 0 \\
\end{array}%
\right).
\end{equation}
Denote $C(t)=(c_1(t),c_2(t),c_3(t))^T$. To control such a three-level
system, we use the control Hamiltonian of
$H_u=u_1(t)H_1+u_2(t)H_2$, where
\begin{equation}\small \label{level3system3}
H_{1}=
\left(%
\begin{array}{ccc}
  0 & 0 & 0 \\
  0 & 0  & 1 \\
  0 & 1  & 0 \\
\end{array}%
\right), \ H_{2}=
\left(%
\begin{array}{ccc}
  0 & 0 & 1 \\
  0 & 0  & 0 \\
  1 & 0  & 0 \\
\end{array}%
\right).
\end{equation}
Similarly, we describe the inhomogeneous three-level quantum
ensemble with the Hamiltonian of each member as
\begin{equation}\small \label{level3system4}
H_{\varepsilon_0,\varepsilon_u}(t)=\varepsilon_0H_0+\varepsilon_u((u_{1}(t)H_1+u_{2}(t)H_2).
\end{equation}
Suppose that the inhomogeneous quantum ensemble consists of three
classes of members labeled with classes \emph{A}, \emph{B} and
\emph{C}, respectively. For this multiclass QEC problem, we first use the same control field to drive the
members belonging to classes \emph{A}, \emph{B} and \emph{C} from an
initial state $|\psi_{0}\rangle$ to three different target
eigenstates ($|1\rangle$, $|2\rangle$ and $|3\rangle$),
respectively, so that we can classify them with an additional
physical operation (e.g., projective measurement).

We can modify \emph{Algorithm 2}
into its multiclass version and then apply it to the three-level
inhomogeneous quantum ensemble for finding an optimal control strategy
$u^*(t)=\{u^*_m(t), m=1,2\}$ to maximize the performance function
\begin{equation}\small \label{level3system5}
\begin{split}
J(u)=&\frac{1}{3}(\mathbb{E}[J^A]+\mathbb{E}[J^B]+\mathbb{E}[J^C])\\
=&\frac{1}{3}(\mathbb{E}[F^2(|\psi^A_{\varepsilon_{0},\varepsilon_{u}}(T)\rangle,|1\rangle)]+\mathbb{E}[F^2(|\psi^B_{\varepsilon_{0},\varepsilon_{u}}(T)\rangle,|2\rangle)]\\
&+\mathbb{E}[F^2(|\psi^C_{\varepsilon_{0},\varepsilon_{u}}(T)\rangle,|3\rangle)]).
\end{split}
\end{equation}

\begin{figure}
\centering
\includegraphics[width=4.5in]{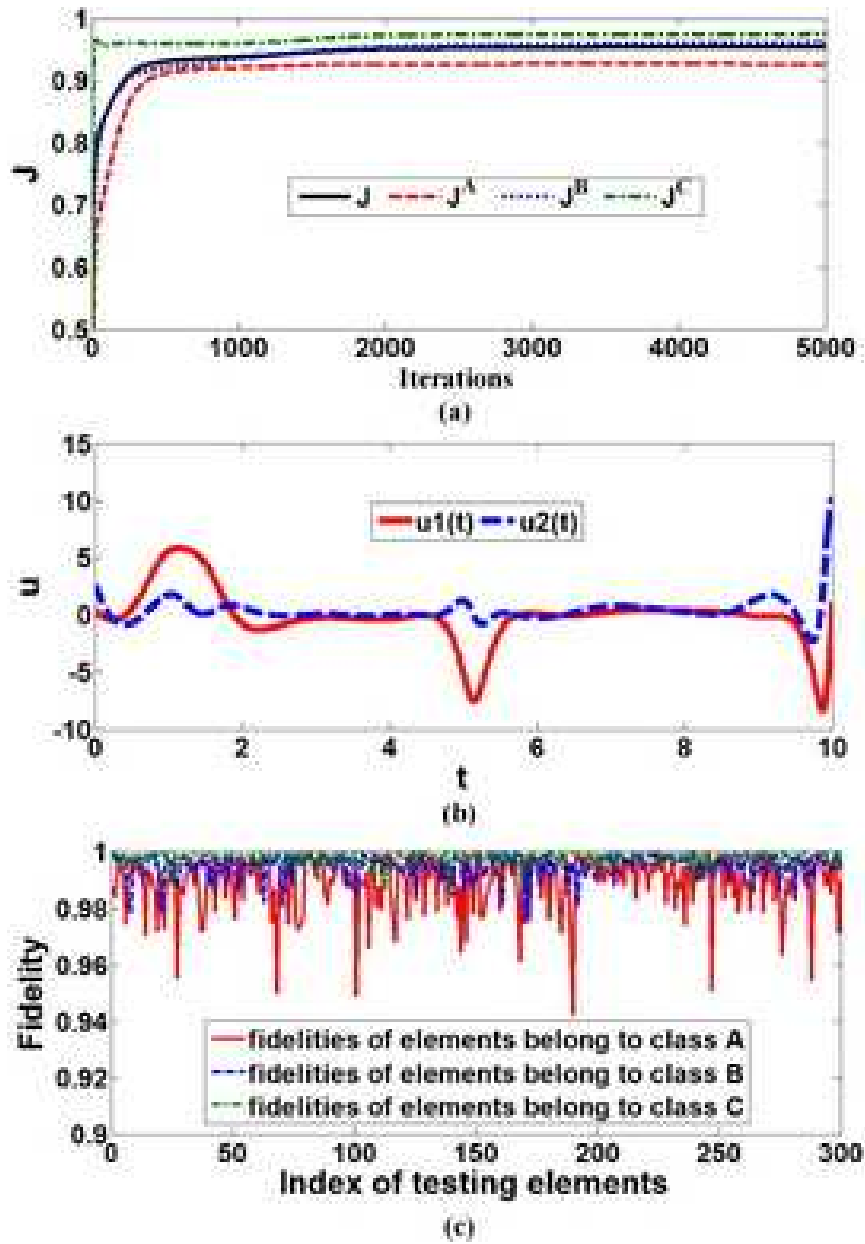}
\caption{Learning control performance for multiclass QEC: (a)
evolution of performance function $J(u)$ with its three component
values $J^A(u)$, $J^B(u)$ and $J^C(u)$; (b) the learned optimal
control for the multiclass QEC problem; (c) control performance
regarding fidelity.}
\end{figure}

The parameter settings are listed as follows: the initial state
$C_0=(\frac{1}{\sqrt{3}},\frac{1}{\sqrt{3}},\frac{1}{\sqrt{3}})$;
the three target eigenstates for classes \emph{A}, \emph{B} and
\emph{C} are $C_{\text{targetA}}=(1,0,0)$ (i.e., $|1\rangle$),
$C_{\text{targetB}}=(0,1,0)$ (i.e., $|2\rangle$) and
$C_{\text{targetA}}=(0,0,1)$ (i.e., $|3\rangle$), respectively;
the ending time $T=10$ (in atomic unit) and the total time
duration $[0,T]$ is equally discretized into $Q=1000$ time slices;
the learning rate is $\eta_k=0.2$; the control is initialized as
$u^{0}(t)=\{u^0_1(t)=\text{sin}t, u^0_2(t)=\text{sin}t\}$.
%; the
%threshhold $\epsilon=10^{-4}$ and if
%$|J(u^{k+1})-J(u^k)|<\epsilon$ for uninterrupted $n_e=100$
%learning steps, we find a satisfactory control for the multiclass
%QEC problem.
The parameters $\varepsilon_{0}$ and
$\varepsilon_{u}$ characterize the inhomogeniety of the quantum
ensemble and they have different normal distributions that are
described with the distribution functions
$d_0(\varepsilon_0)=\Phi(\frac{\varepsilon_{0}-\mu_0}{\sigma_0})$
and
$d_u(\varepsilon_u)=\Phi(\frac{\varepsilon_{u}-\mu_u}{\sigma_u})$,
where for class \emph{A} $(\mu_0^A=1,3\sigma_0^A=0.05)$ and
$(\mu_u^A=0.8,3\sigma_u^A=0.05)$, for class \emph{B}
$(\mu_0^B=0.8,3\sigma_0^B=0.05)$ and
$(\mu_u^B=1,3\sigma_u^B=0.05)$, for class \emph{C}
$(\mu_0^C=1.2,3\sigma_0^C=0.05)$ and
$(\mu_u^C=1.2,3\sigma_u^C=0.05)$. To construct the generalized
system for learning the optimal control, the sampling method as
described in \eqref{discreteA} is adopted with setting
$N^A_{\varepsilon_0}=N^A_{\varepsilon_u}=N^B_{\varepsilon_0}=N^B_{\varepsilon_u}=N^C_{\varepsilon_0}=N^C_{\varepsilon_u}=3$.

The learning performance is shown in Fig. 15. The evolution of the
performance function $J(u)$ is shown in Fig. 15(a) with its three
component values $J^A(u)$, $J^B(u)$ and $J^C(u)$. The results demonstrate that the SLC based classification method is
effective for
multiclass QEC of multi-level quantum ensembles and has good scalability. The learned optimized
control for the coherent control step of QEC is shown in Fig. 15(b).
As shown in Fig. 15(c), $300$ randomly selected samples for each
class of members are tested and all of them are
controlled to their corresponding target eigenstates, respectively, with high
fidelity. The mean value of fidelity for testing of \emph{class A}
is 0.9897, for \emph{class B} is 0.9953 and for \emph{class C} is
0.9976. With additional $10^4$ testing samples for each class, we
have the classification accuracy $\zeta = 98.80\%$, which
verifies the effectiveness of the proposed SLC based approach for
QEC.

\section{Conclusions}\label{Sec6}
In this paper, we present a systematic classification approach for
inhomogeneous quantum ensembles by combining an SLC approach with quantum
discrimination. The classification process is
accomplished via simultaneously steering members belonging to different classes to different corresponding target states
(e.g., eigenstates). A new discrimination method is first presented for
quantum systems with similar Hamiltonians. Then an SLC method is
proposed for quantum ensemble classification.
Numerical experiments are carried out to test the
performance of the proposed approach for the binary
classification of  two-level quantum ensembles and the
multiclass classification of three-level $\Lambda$-type quantum ensembles. All the numerical results demonstrate the effectiveness of the proposed approach for quantum ensemble classification.

\appendix
Recall that $J^a_{\varepsilon^a_0,\varepsilon^a_u}(u)=\vert
\langle\psi^a_{\varepsilon^a_0,\varepsilon^a_u}(T)\vert\psi_{\textrm{targetA}}\rangle\vert^2$
and $\vert\psi^a_{\varepsilon^a_0,\varepsilon^a_u}(t)\rangle$
satisfies
\begin{equation}\label{app-eq:sch}
\frac{d}{dt}\vert\psi^a_{\varepsilon^a_0,\varepsilon^a_u}(t)\rangle=-iH^a_{\varepsilon^a_0,\varepsilon^a_u}(t)\vert\psi^a_{\varepsilon^a_0,\varepsilon^a_u}(t)\rangle,\quad
\vert\psi^a_{\varepsilon^a_0,\varepsilon^a_u}(0)\rangle=\vert\psi_{0}\rangle.
\end{equation}
For ease of notation, we consider the case where only one
control is involved, i.e.,
$H^a_{\varepsilon^a_0,\varepsilon^a_u}(t)=g_0(\varepsilon^a_0)H_0+u(t)g_u(\varepsilon^a_u)H_1$.
The expression of the gradient of
$J^a_{\varepsilon^a_0,\varepsilon^a_u}(u)$ with respect to the
control $u$ can be derived by using a first order perturbation.

Let $\delta\psi(t)$ be the modification of $\vert \psi(t)\rangle$
induced by a perturbation of the control from $u(t)$ to
$u(t)+\delta u(t)$. By keeping only the first order terms,
we obtain the equation satisfied by $\delta\psi$:
\begin{equation}
\left \{
\begin{split}
\frac{d}{dt}\delta\psi=&-i\left(g_0(\varepsilon^a_0)H_0+u(t)g_u(\varepsilon^a_u)H_1\right)\delta\psi\\
&-i\delta
u(t)g_u(\varepsilon^a_u)H_1\vert\psi^a_{\varepsilon^a_0,\varepsilon^a_u}(t)\rangle,\\
\delta\psi(0)=&0.
\end{split}
\right.
\end{equation}

Let $U_{\varepsilon^a_0,\varepsilon^a_u}(t)$ be the propagator
corresponding to \eqref{app-eq:sch}. Then,
$U_{\varepsilon^a_0,\varepsilon^a_u}(t)$ satisfies
\begin{equation}
\frac{d}{dt}U_{\varepsilon^a_0,\varepsilon^a_u}(t)=-iH^a_{\varepsilon^a_0,\varepsilon^a_u}(t)U_{\varepsilon^a_0,\varepsilon^a_u}(t),\quad
U(0)=I.
\end{equation}
Therefore,
\begin{equation}\label{app-eq:deltapsi}\small
\begin{split}
\delta\psi(T)&=-iU_{\varepsilon^a_0,\varepsilon^a_u}(T)\int_0^T\delta u(t)U^{\dagger}_{\varepsilon^a_0,\varepsilon^a_u}(t)g_u(\varepsilon^a_u)H_1\vert\psi^a_{\varepsilon^a_0,\varepsilon^a_u}(t)\rangle dt\\
&=-iU_{\varepsilon^a_0,\varepsilon^a_u}(T)\int_0^TU^{\dagger}_{\varepsilon^a_0,\varepsilon^a_u}(t)g_u(\varepsilon^a_u)H_1U_{\varepsilon^a_0,\varepsilon^a_u}(t)\delta
u(t)dt~\vert\psi_0\rangle
\end{split}
\end{equation}
where $U^{\dagger}$ is the adjoint of $U$. Using
(\ref{app-eq:deltapsi}), we have
\begin{equation}\label{app-eq:dJ}\footnotesize
\begin{split}
&J^a_{\varepsilon^a_0,\varepsilon^a_u}(u+\delta u)\\
\approx & J^a_{\varepsilon^a_0,\varepsilon^a_u}(u)+2\Re\left(\langle\psi^a_{\varepsilon^a_0,\varepsilon^a_u}(T)\vert\psi_{\textrm{targetA}}\rangle\langle\psi_{\textrm{targetA}}\vert\delta\psi(T)\right)\\
=&J^a_{\varepsilon^a_0,\varepsilon^a_u}(u)+2\Re\left(-i\langle\psi^a_{\varepsilon^a_0,\varepsilon^a_u}(T)\vert\psi_{\textrm{targetA}}\rangle\langle\psi_{\textrm{targetA}}\vert
\int_0^TG^a_1(t)\delta u(t)dt~ \vert\psi_0\rangle\right)\\
=&J^a_{\varepsilon^a_0,\varepsilon^a_u}(u)+\int_0^T2\Im\left(\langle\psi^a_{\varepsilon^a_0,\varepsilon^a_u}(T)\vert\psi_{\textrm{targetA}}\rangle\langle\psi_{\textrm{targetA}}\vert
G^a_1(t)\vert\psi_0\rangle\right)\delta u(t)dt
\end{split}
\end{equation}
where
$G^a_1(t)=U_{\varepsilon^a_0,\varepsilon^a_u}(T)U_{\varepsilon^a_0,\varepsilon^a_u}^\dagger(t)g_u(\varepsilon^a_u)H_1U_{\varepsilon^a_0,\varepsilon^a_u}(t)$,
$\Re(\cdot)$ and $\Im(\cdot)$ denote respectively the real and
imaginary parts of a complex number.

Recall also that the definition of the gradient implies
\begin{equation}\label{app-eq:dJ2}\small
\begin{split}
J^a_{\varepsilon^a_0,\varepsilon^a_u}(u+\delta u)&=J^a_{\varepsilon^a_0,\varepsilon^a_u}(u)+\langle \nabla J^a_{\varepsilon^a_0,\varepsilon^a_u}(u),\delta u\rangle_{L^2([0,T])}+o(\Vert\delta u\Vert)\\
&=J^a_{\varepsilon^a_0,\varepsilon^a_u}(u)+\int_0^T \nabla
J^a_{\varepsilon^a_0,\varepsilon^a_u}(u)\delta
u(t)dt+o(\Vert\delta u\Vert).
\end{split}
\end{equation}
Therefore, by identifying \eqref{app-eq:dJ} with \eqref{app-eq:dJ2}, we obtain
\begin{equation}\label{app-eq:gradJ}
 \nabla J^a_{\varepsilon^a_0,\varepsilon^a_u}(u)=2\Im\left(\langle\psi^a_{\varepsilon^a_0,\varepsilon^a_u}(T)\vert\psi_{\textrm{targetA}}\rangle\langle\psi_{\textrm{targetA}}\vert G^a_1(t)\vert\psi_0\rangle\right).
\end{equation}

%%%%%%%%%%%%%%%%%%%%%%%%%%%%%%%%%%%%%%%%%%%%%%%%%%%%%%%%%%%%%%%%%%%%%%%%%%%%%%%%

\end{document}